\documentclass[10pt,twocolumn,twoside,journal]{IEEEtran} 
\usepackage{cite}
\usepackage{amsmath} 
\usepackage{amssymb}  
\usepackage{amsfonts}
\usepackage{mathrsfs}
\usepackage{mathtools}
\usepackage[thmmarks, amsmath]{ntheorem}
\usepackage{booktabs}
\usepackage[bbgreekl]{mathbbol}
\usepackage[dvipsnames,table]{xcolor}
\usepackage[cal=cm]{mathalfa}
\DeclareSymbolFontAlphabet{\mathbb}{AMSb}
\DeclareSymbolFontAlphabet{\mathbbl}{bbold}
\DeclareFontEncoding{LS1}{}{}
\DeclareFontSubstitution{LS1}{stix}{m}{n}
\DeclareSymbolFont{stixsymbols}       {LS1}{stixscr}  {m} {n}
\DeclareSymbolFontAlphabet{\mathscrl} {stixsymbols}

\DeclareMathOperator{\diff}{d}
\newcommand{\ddt}{\tfrac{\diff}{\diff \!t}}
\newtheorem{theorem}{Theorem}
\newtheorem{definition}{Definition}
\newtheorem{remark}{Remark}
\newcommand{\round}[1]{\ensuremath{\lfloor#1\rceil}}

\begin{document}
\title{Dual-port grid-forming control of MMCs and its applications to grids of grids
}
      
\author{Dominic~Gro\ss{},~\IEEEmembership{Member,~IEEE,}
        Enric S\'anchez-S\'anchez,~\IEEEmembership{Member,~IEEE,}
        Eduardo Prieto-Araujo,~\IEEEmembership{Senior Member,~IEEE,}
	and~Oriol Gomis-Bellmunt,~\IEEEmembership{Fellow,~IEEE}

\thanks{The work of Oriol Gomis-Bellmunt is supported by the ICREA Academia program. Eduardo Prieto-Araujo is a Serra H\'unter Lecturer. This work was also funded by FEDER / Ministerio de Ciencia, Innovaci\'on y Universidades - Agencia Estatal de Investigaci\'on, Project RTI2018-095429-B-I00. 

D. Gro\ss{} is with the Department of Electrical and Computer Engineering at the University of Wisconsin-Madison, Madison, USA. E. Prieto-Araujo, and O. Gomis-Bellmunt are with and E. S\'anchez-S\'anchez was with the Centre d'Innovaci\'o Tecnol\`ogica en Convertidors Est\`atics i Accionaments, Departament d'Enginyeria El\`ectrica, Universitat Polit\`ecnica de Catalunya, Barcelona, Spain; e-mail: dominic.gross@wisc.edu, enric.sanchez.sanchez@gmail.com; eduardo.prieto-araujo@citcea.upc.edu; oriol.gomis@upc.edu}
}

\maketitle
\begin{abstract}
This work focuses on grid-forming (GFM) control of Interconnecting Power Converters (IPCs) that are used to interconnect multiple HVAC and HVDC subgrids to form a grid of grids. We introduce the concept of dual-port GFM control that leverages the ability of Modular Multilevel Converters (MMCs) to simultaneously form its AC and DC terminal voltage and present two dual-port GFM MMC controls. We provide analytical results and high-fidelity simulations that demonstrate that (i) dual-port GFM control is more resilient to contingencies (i.e., line and generator outages) than state-of-the-art single-port GFM control, and (ii) unlike single-port GFM control, dual-port GFM control does not require assigning grid-forming and grid-following (GFL) roles to the IPC terminals in grids of grids. Finally, we provide an in-depth discussion and comparison of single-port GFM control and the proposed dual-port GFM controls.
\end{abstract}

% \begin{IEEEkeywords}
% some keywords
% \end{IEEEkeywords}

\section{Introduction}
A major transition in the operation of electric power systems is the increasing integration of power electronic converters that interface renewable generation, energy storage systems, high voltage direct current (HVDC) transmission, and industrial and domestic loads. Replacing synchronous generators with converter-interfaced resources results in significantly different power system dynamics and challenges standard operating paradigms. In particular, while power converters have limited inertia and reduced overload capability, they are fully controllable and enable a fast and flexible response as long as their limitations are considered \cite{LCP20,LMI16,PBE+21}. 

The use of power electronic converters in HVDC transmission systems has resulted in the emergence of segmented power systems composed of multiple HVAC subgrids interconnected by means of point-to-point HVDC links. The proliferation of HVDC grids using Voltage Source Converters (VSC) will enable more complex interconnections of multiple meshed HVAC and HVDC subgrids through Interconnecting Power Converters (IPC).

Typically, control strategies for DC/AC VSCs are broadly categorized into (i) grid-following (GFL) controls that assume a stable AC voltage (i.e., magnitude and frequency) and (ii) grid-forming (GFM) strategies that form a stable AC voltage (i.e., magnitude and frequency) at the converter terminal. As a consequence of relying on a stable AC voltage, GFL control may fail due to voltage disturbances \cite{NERC17} or if insufficient GFM units (i.e., synchronous generators or GFM converters) are online to ensure frequency stability.

In contrast, GFM power converters can form a stable grid and are envisioned to be the cornerstone of future power systems. The prevalent approaches to GFM control are so-called droop-control \cite{MCC-DMD-RA:93}, synchronous machine emulation~\cite{DSF15}, and (dispatchable) virtual oscillator control \cite{JD+2014,GCB+19}. All of the aforementioned controls form a stable AC voltage waveform and provide primary frequency control. However, they require a stable DC voltage and will destabilize the system if the DC voltage is not tightly controlled \cite{TGA+20}. 

On the other hand, in the context of HVDC systems, VSC controls have been proposed that stabilize the DC voltage but require a stable AC voltage (i.e., frequency and magnitude) and will destabilize the DC system if the AC voltage is not tightly controlled \cite{GSA+20}. Consequently, GFM controls can be broadly categorized into AC grid-forming (AC-GFM) and DC grid-forming (DC-GFM). {In the existing literature it is commonly assumed that AC-GFM and DC-GFM are mutually exclusive concepts.} Therefore, operating such a system with standard AC-GFM and DC-GFM controls requires assigning AC-GFM and DC-GFM roles to different IPCs to ensure stability of the individual HVAC and HVDC subgrids \cite{GSA+20}. This task is non-trivial and can result in a system with complex dynamics that is vulnerable to changes in the subsystem topologies or control reserves (e.g., due to contingencies). Our key contribution is the concept of dual-port GFM control that does not require assigning AC-GFM or DC-GFM roles to different IPCs but uses the same control on all IPCs. 

Today Modular Multilevel Converters (MMC) are emerging as the key Interconnecting Power Converter (IPC) technology that allows to interconnect different high-voltage AC and DC subgrids. A key feature of the MMC is that it can directly control both its AC and DC terminal voltages \cite{LV14} and leverage the energy stored in its arms' cells to provide limited inherent energy storage functionalities. This degree of freedom can be used to enhance the converter and overall system performance \cite{SGP+20,FAR+19}. To the best of our knowledge, GFM MMC controls available in the literature are single-port GFM, i.e., either AC-GFM or DC-GFM, {and control} the MMC's internal energy through GFL control on the other terminal (AC-GFL/DC-GFM or AC-GFM/DC-GFL). {A notable exception is the control proposed in \cite{MSD+2020} that can control the MMC's internal energy simultaneously through its AC and DC terminal. While this control provides DC voltage control and partial AC-GFM features it generally requires a stable AC voltage.}

The main contribution of this paper is the novel concept of dual-port GFM control that leverages the MMC's degrees of freedom to simultaneously form the voltage on the MMC's AC and DC terminal and control the MMC energy through both GFM terminals. In contrast to single-port GFM control, dual-port GFM control (i) allows operating a grid of grids with only one control concept and overcomes the need for assigning roles to different IPCs, and (ii) survives line and generator outages on either the AC or DC network. 

We propose two dual-port GFM MMC controls, present guidelines for selecting the dual-port GFM control parameters, and provide small-signal stability results for a simplified test system that highlight the main features of the dual-port GFM controls. Finally, we use high-fidelity case studies to compare single-port and dual-port GFM controls, highlight the increased resilience of dual-port GFM control, and illustrate the use of dual-port GFM controls in a grid of grids.

This manuscript is organized as follows. Sec.~\ref{sec:setup} introduces our problem setup, objectives, and terminology. Sec.~\ref{sec:MMC} reviews standard MMC models and  MMC controls used as throughout this manuscript. In Sec.~\ref{sec:sosgridform} we review standard (single-port) GFM MMC controls and critically reflect on their properties. Our main contribution are the two novel dual-port GFM controls introduced in Sec.~\ref{sec:dualgridform}. Analytical small-signal stability results for a simplified setup are presented in Sec.~\ref{sec:stability}. In Sec.~\ref{sec:case_study} detailed simulation studies are used to validate the analytical results, compare the novel dual-port GFM controls to standard single-port GFM controls, and illustrate operation of a grid of grids with dual-port GFM control. Sec.~\ref{sec:disc} discusses and compares the proposed controls and state of the art and Sec.~\ref{sec:conclusions} provides the conclusions and discusses future research directions.

\section{Motivation and problem setup}\label{sec:setup}
To introduce and motivate the problem setup considered in this work, this section briefly reviews the concept of grids of grids and main control objectives of IPC control.

\subsection{The grid of grids}\label{sec:gogs}
 Multiple AC subgrids can be interconnected by means of HVDC transmission to e.g., integrate offshore wind, exchange power between different networks to increase reliability and flexibility, and transport electric power over long distances or through undersea cables (please see \cite{GSA+20} for a detailed discussion). Moreover, one may choose to segment existing AC systems into different AC and DC subgrids and leverage the additional controllability of HVDC transmission to prevent disturbances and faults from spreading across the grid of grids. IPCs are responsible to ensure a stable exchange of power between the different interconnected sub-systems, while also ensuring the overall system stability whenever a contingency occurs. In particular, IPCs can prevent faults and disturbances from spreading and provide frequency and voltage support from neighbouring sub-systems.

\begin{figure}[b!!]
	\centering
	\includegraphics[width=1\columnwidth]{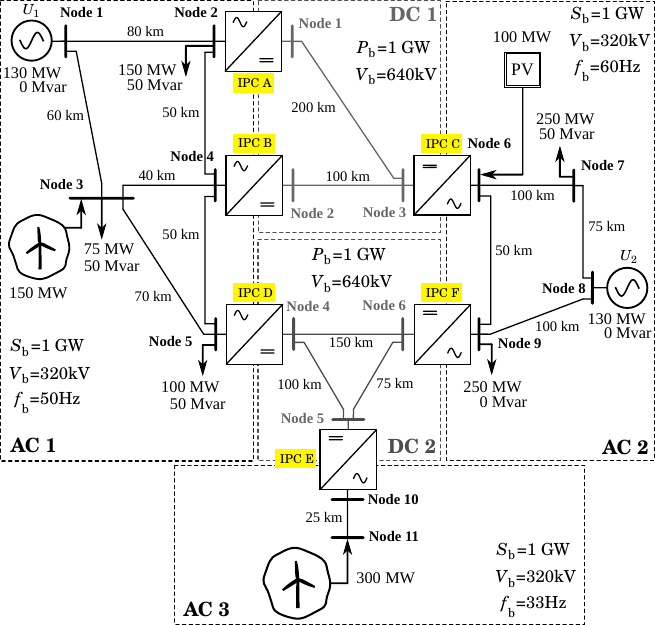}
	\caption{Example for a grid containing multiple AC and DC subgrids \cite{GSA+20}.\label{fig:gogs}}
\end{figure}

Grids of grids are rapidly emerging in practice. For instance, the European system interconnects several (non-synchronous) AC subgrids (i.e. continental Europe, Great Britain, Ireland, Nordic system), through an ever increasing number of HVDC connections. Moreover, multi-terminal HVDC systems are already in operation in China. The widespread adoption of HVDC technology will likely result in more complex systems combining multiple meshed AC and DC subgrids similar to the example of a grid of grids shown in Fig.~\ref{fig:gogs}. 

\begin{figure}
\centering
	\includegraphics[width=.55\columnwidth]{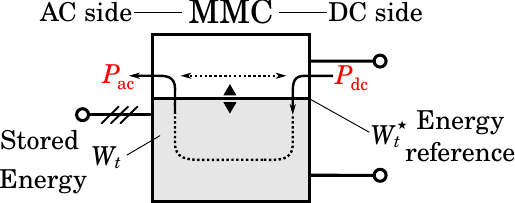}
	\caption{Interpretation of the MMC internal as energy buffer. \label{fig:watertank}}
\end{figure}

Operating a future grid of grids will require different IPCs to form different AC and DC grids. Using state-of-the-art MMC controls, converters providing grid-forming functionalities on their AC (or DC) terminals will contribute to stabilizing the AC (or DC) network but act as controlled power source / load on the DC (or AC) network to control their internal energy (see Fig.~\ref{fig:watertank}) and fail if the DC (or AC) terminal is not stabilized by another device. This raises the question of how to decide which IPC should support which grid \cite{GSA+20} to achieve stable and resilient operation of a grid of grids.
The main contribution of this work is to resolve these challenges using a novel control concept for MMCs that uses the same control on all IPCs, i.e., does not require assigning different controls to different IPCs.

\subsection{IPC control objectives and definitions}\label{sec:setpoints}
The nominal operating point $(\omega^\star,P^\star_{\text{ac}},P^\star_{\text{dc}},V^{\text{dc}\star}_t,V^{\text{ac}\star}_t)$ of the IPCs is typically defined by a system-level control that aims to minimize generation costs subject to system constraints (e.g., thermal and voltage limits), IPC constraints (e.g., current limits), and possibly security (e.g., $N-1$) constraints. In a grid of grids setting, the nominal operating point can be computed by periodically (i.e., every few minutes or hours) solving an OPF problem that contains the network power flow equations for the AC and DC subgrids, and the power balance constraint $P^\star_{\text{ac}} = P^\star_{\text{dc}} -P_{\text{loss}}$ for IPC nodes. Please see \cite[Sec. V]{GSA+20} for further details. For clarity of the presentation, we assume that IPCs are lossless, i.e., $P_{\text{loss}}=0$.

{
The IPC control should meet three key control objectives. 
\begin{enumerate}
 \item During nominal steady-state operation the IPC needs to be controlled to the nominal operating point (i.e., AC voltage, DC voltage, and power transfer) periodically prescribed by a system-level controller.
 \item The IPC should autonomously respond to variations in load or generation and contingencies to stabilize the grid of grids until a system-level controller can identify the new situation and provide an updated nominal steady-state operating point that is consistent with the post-contingency system configuration.
 \item The IPC needs to autonomously adjust its AC and/or DC power injection to stabilize its internal energy storage elements (see Sec.~\ref{sec:MMC}).
\end{enumerate}
We emphasize that the time scales of internal energy control, converter-based primary frequency control in AC networks and DC voltage control in HVDC systems partially overlap. As a consequence, all of these signals need to be controlled on the same time-scales. Doing so is a significant task that is not fully addressed by existing controls (see Sec.~\ref{sec:sosgridform:disc}). In contrast, power flow control through infrequent setpoint updates is inherently associated with longer time scales.} To broadly categorize IPC controls, we use the following definition.

\begin{definition}{\bf{(Grid-forming and grid-following control)}}\label{def:GFM}
 For each converter terminal (i.e., AC and DC), we refer to a control as grid-forming (GFM) if it forms a stable voltage at the converter terminal (i.e., AC-GFM or DC-GFM), and grid-following (GFL) if it requires another device to stabilize the voltage at the converter terminal (i.e., AC-GFL or DC-GFL). 
\end{definition}
We emphasize that there is no precise and agreed upon definition of GFM and GFL for either AC or DC terminals and resolving this challenge is beyond the scope of this manuscript. Instead, our definition highlights the main salient feature of GFM control. The GFM functionality becomes evident when disconnecting the IPC from the AC or DC network. While a GFL control cannot maintain stability in this scenario, GFM control ensures that voltage is controlled. For example, DC-GFM refers to the IPCs ability to control DC voltage without relying on other sources. This can be achieved by a DC voltage controller or enforcing a droop characteristic between voltage and power (or current).

{Notably, Definition \ref{def:GFM} does not specify control structures, but rather specifies a functional requirement. For example, the MMC control in \cite{MSD+2020} uses $P_\text{ac}$-$f$ droop to compute the phase angle of the MMC AC voltage. However, the AC active power setpoint for the $P_\text{ac}$-$f$ droop control is provided by a proportional-integral (PI) DC voltage control. An MMC with this control is stable if its AC terminal is connected to an AC grid with frequency control and its DC terminal is connected to a constant power load (see \cite[Fig. 2]{MSD+2020}). However, it is not stable if the DC terminal is connected to a DC voltage source with $P_\text{dc}$-$V^\text{dc}_t$ droop and the AC terminal is connected to a constant power load (i.e., Fig.~\ref{fig:simptest} with $s_1$ open, $s_2$ closed). Therefore, according to Definition \ref{def:GFM}, the control in \cite{MSD+2020} is AC-GFL/DC-GFM.} 

To the best of our knowledge, GFM MMC controls in the literature are either AC-GFM or DC-GFM while controlling the MMC's internal energy through GFL control on the opposite terminal (e.g., AC-GFM/DC-GFL or AC-GFL/DC-GFM). We will refer to these standard controls as \emph{single-port GFM}. {A notable exception is the AC-GFL/DC-GFM control proposed in \cite{MSD+2020} that can control the MMC's internal energy simultaneously through both its AC and DC terminal.}

The main contribution of this work are \emph{dual-port GFM} controls that provide GFM functions on both the AC and DC terminal and simultaneously control the MMC's internal energy through both GFM terminals.

\section{MMC model and basic control}\label{sec:MMC}
This reviews standard MMC models and basic internal MMC controls that form the basis for GFM MMC controls and will be used in our case studies.
\subsection{Electrical MMC model}\label{sec:MMC:electrical}
\begin{figure}[b!!!]
	\centering
	\includegraphics[width=1\columnwidth]{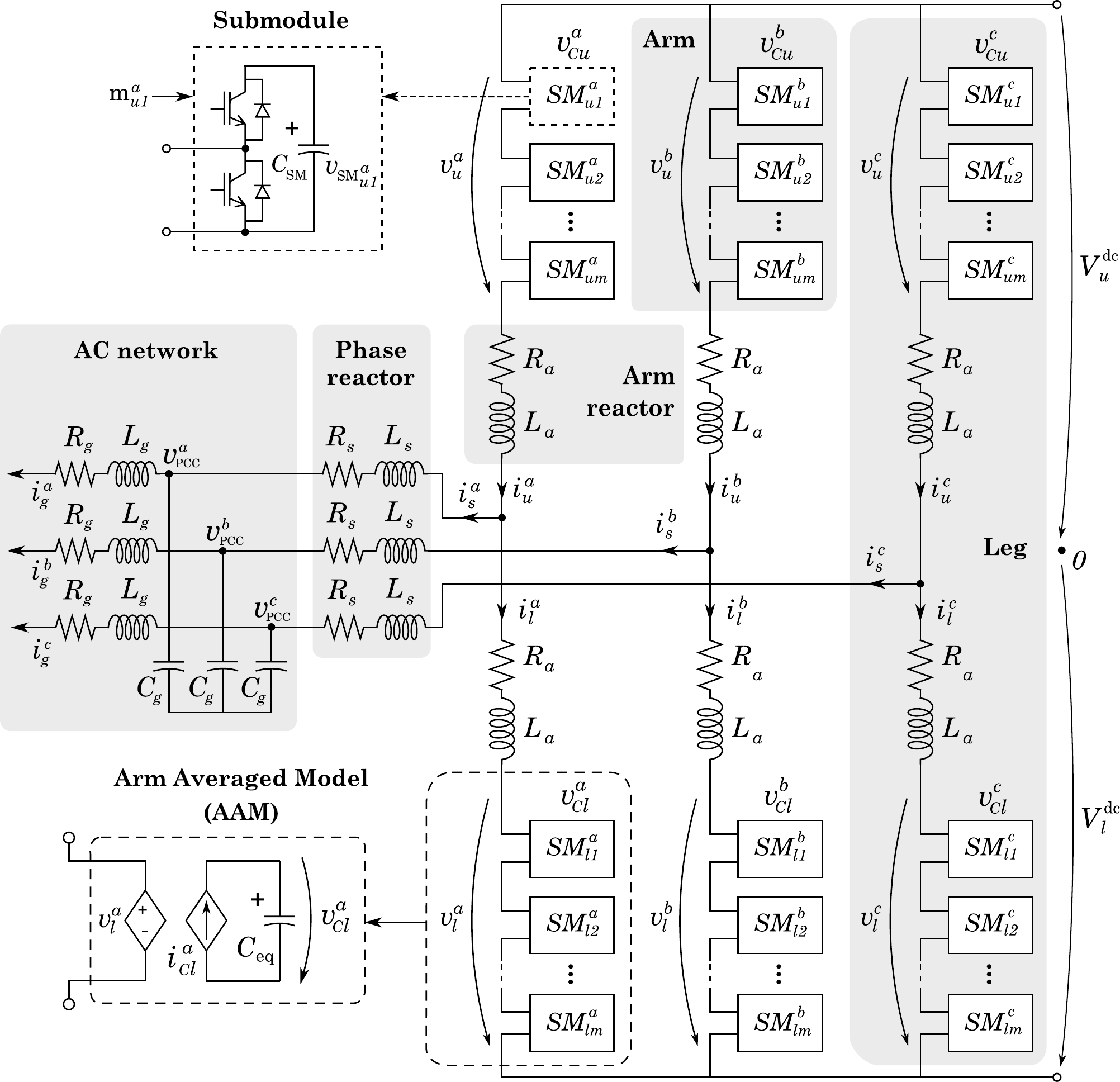}
	\caption{Electrical model of MMC connected to an AC network (including lines or cables) with load or generation. \label{fig:mmc_circuit}}
\end{figure}

The MMC consists of six arms. Each arm contains $N_{\text{arm}}$ submodules that are connected in series with an arm reactor (see Fig.~\ref{fig:mmc_circuit}). Each individual submodule is controlled to either insert or bypass its capacitor through modulation signals $m^j_{u,k}$ and $m^j_{l,k}$, where $j\in \{a, b, c\}$ denotes the three phases that are also referred to as legs and contain an upper and lower arm. Each arm is controlled to synthesize the voltage required to achieve the desired power exchange between the AC and the DC sides and internally balance the energy stored in its converter submodules (see e.g. \cite{QZL11}).
Following the conventions in~\cite{Prieto-Araujo2017a}, we use the \emph{diff-sum} coordinates widely used in the literature on control and modeling of MMCs, i.e., $v_{\text{diff}}^j \coloneqq \tfrac{1}{2}(-v_{u}^j+v_{l}^j)$ denotes the differential voltage (middle point of the arm), $v_{\text{sum}}^j \coloneqq v_{u}^{j}+v_{l}^{j}$ and $i_{\text{sum}}^j \coloneqq \tfrac{1}{2}(i_{u}^j+i_{l}^j)$ are the additive voltage (approx. equal to the DC voltage) and additive current (from upper to lower arm). Moreover, $R_a$ and $L_a$ denote arm resistance and inductance, $R_s$ and $L_s$ the AC filter resistance and inductance, and $R_g$, $L_g$, and $C_g$ are the AC network resistance, inductance, and capacitance. Assuming a balanced DC and AC side and grounded AC neutral point, the MMC current dynamics are modeled by
\begin{subequations}
\begin{align}
L_{\text{eq}} I_3\ddt i_s^{abc} &= -R_{\text{eq}} I_3i_s^{abc}+v_{\text{diff}}^{abc}-v^{abc}_{\text{PCC}} \label{eq:mmc1_final}\\
C_{\text{ac}}I_3\ddt u^{abc}  &= i_s^{abc}-i_g^{abc} \label{eq:mmc2_final} \\
2L_aI_3\ddt i_{\text{sum}}^{abc} &= -2R_aI_3i_{\text{sum}}^{abc} + \mathbbl{1}_3 V_t^{\text{dc}}  -v_{\text{sum}}^{abc} \label{eq:mmc3_final},
\end{align}
\end{subequations}
where $v^{abc}_{\text{PCC}} \in \mathbb{R}^3$ denotes the PCC voltage, $i_s^{abc} \in \mathbb{R}^3$ is the AC grid current, i.e., $i_s^j=i_u^j-i_l^j$ for $j \in \{a,b,c\}$, and $V_t^{\text{dc}} \in \mathbb{R}$ is the DC side voltage. Moreover, $R_{\text{eq}}\coloneqq R_s+\tfrac{R_a}{2}$ and $L_{\text{eq}}\coloneqq L_s+\tfrac{L_a}{2}$, and $I_n$ is the $n\times n$ identity matrix, and $\mathbbl{1}_n$ denotes a vector of ones of length $n$. Using the Average Arm Model (AAM)~\cite{Harnefors2013,Saad2015}, the equivalent capacitor voltage $v_{Cu}^j=\sum_{k=1}^{N_\text{arm}} v_{\text{\tiny{SM}}_{u,k}^j} \in \mathbb{R}$ and $v_{Cl}^j =\sum_{k=1}^{N_\text{arm}} v_{\text{\tiny{SM}}_{l,k}^j}\in \mathbb{R}$ of each arm depends on the power exchanged by the arm and the equivalent capacitance $C_{\text{eq}}$
\begin{align}\label{eq:ceq}
\tfrac{1}{2} C_{\text{eq}} \ddt (v_{Cu}^j)^2 = v^j_u i^j_u, \quad \tfrac{1}{2} C_{\text{eq}} \ddt (v_{Cl}^j)^2 = v^j_l i^j_l.
\end{align}
We will use the Average Arm Model (AAM) for control design and analysis, while the model shown in Fig.~\ref{fig:mmc_circuit} with individual submodule capacitor charge dynamics and ideal switches will be used for simulation studies that validate and illustrate the results.

\subsection{Basic internal MMC controls}
\begin{figure}[htbp!]
	\centering
	\includegraphics[width=1\columnwidth]{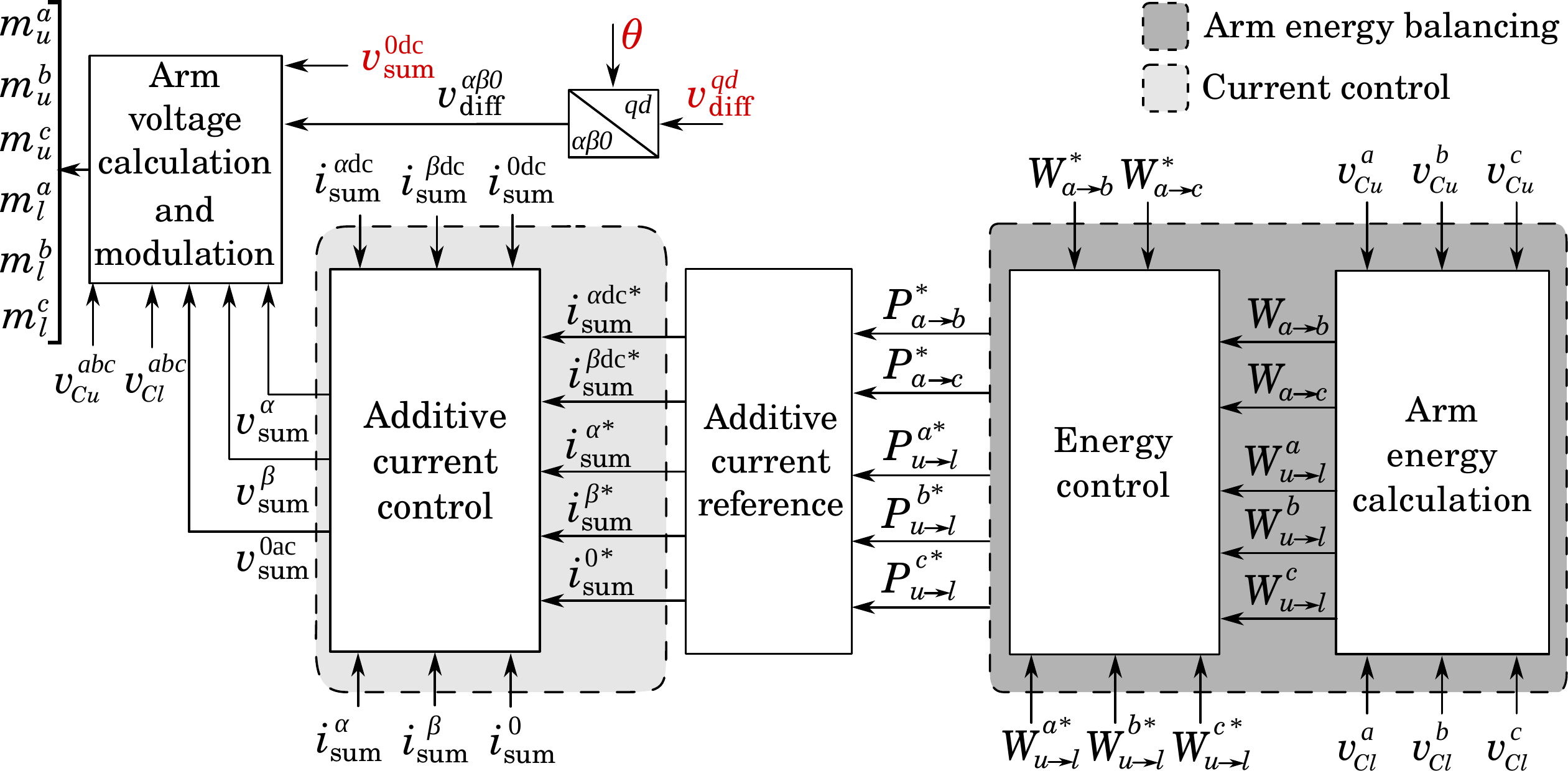}
	\caption{MMC internal energy-balancing control architecture with AC and DC terminal control inputs (red).  \label{fig:mmc_internalbalcontrol}}
	\vspace{1em}
		\includegraphics[width=1\columnwidth]{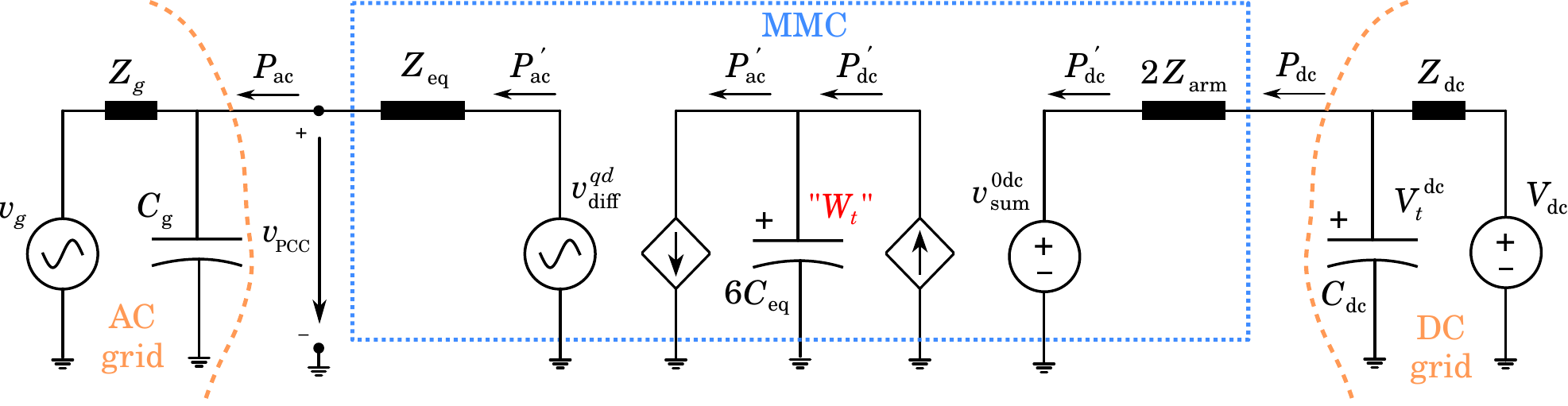}
	\caption{Macroscopic system-level model of an MMC in closed-loop with the internal energy-balancing control shown in Fig.~\ref{fig:mmc_internalbalcontrol}. \label{fig:mmcabstracten}}
\end{figure}

In the present study, we use an energy-based control approach to explicitly control the internal energy of the MMC in closed-loop. The basic MMC internal energy-balancing control structure is shown in Fig.~\ref{fig:mmc_internalbalcontrol} and includes five PI controllers that balance the energy among the legs (horizontal balancing) and between upper and lower arms (vertical balancing) through inner PI current controllers that control MMC's additive currents \cite{Prieto-Araujo2017a}. We emphasize that the AC and DC components of the MMC's internal currents play specific roles in the energy transfer within the MMC (see \cite{Prieto-Araujo2017a} for further details). In the remainder, we assume that the control shown in Fig.~\ref{fig:mmc_internalbalcontrol} ensures fast and accurate horizontal and vertical energy balancing. This allows us to consider the simplified model shown in Fig.~\ref{fig:mmcabstracten} in which an equivalent capacitor represents the total internal energy 
\begin{align}
W_t=\sum\nolimits_{j\in\{a,b,c\}}\tfrac{1}{2} C_{\text{eq}} ((v^j_{Cu})^2+(v^j_{Cl})^2) 
\end{align}
of the MMC.
Moreover, the vector AC current control in a synchronous $qd0$ reference frame (i.e., using electrical machine notation) with reference angle $\theta$ (provided by an AC-GFM control or phase-locked loop) and the DC current control shown in Fig.~\ref{fig:mmc_currcontrol} control the MMC's AC and DC current. We emphasize that the horizontal and vertical balancing controller and inner DC and AC current controls are independent of the MMC's mode of operation (e.g., AC-GFM, AC-GFL) and are continuously used in all configurations to ensure stability of MMC's internal currents and capacitor voltages. 

To this end, the inverse Clarke transformation is applied to the voltage references for $v^{\alpha\beta 0}_{\text{sum}}$ and $v^{\alpha\beta 0}_{\text{diff}}$ prescribed by the additive current control and outer control to obtain reference values for $v^j_{\text{sum}}$ and $v^j_{\text{diff}}$. Next, the change of coordinates presented in Sec.~\ref{sec:MMC:electrical} is used to obtain reference values  $v^j_{u}$ and $v^j_{l}$ in the original coordinates. Finally, different modulation schemes can be used to obtain the modulation signals $m^j_{u,k}$ and $m^j_{l,k}$. Typically, the number of submodules to be inserted by each arm is computed using $N^j_u=\round{v^j_{u} / v^j_{Cu}}$ and $N^j_l=\round{v^j_{l} / v^j_{Cl}}$, where $\round{\cdot}$ denotes rounding to the nearest integer, and depending on the sign of the arm currents submodules are inserted in ascending or descending order of their voltage to achieve voltage balancing within each arm (see e.g. \cite{QZL11}).

\subsection{Macroscopic MMC model for system-level control design}
\begin{figure}[htbp]
    \centering
    \includegraphics[width=1\columnwidth]{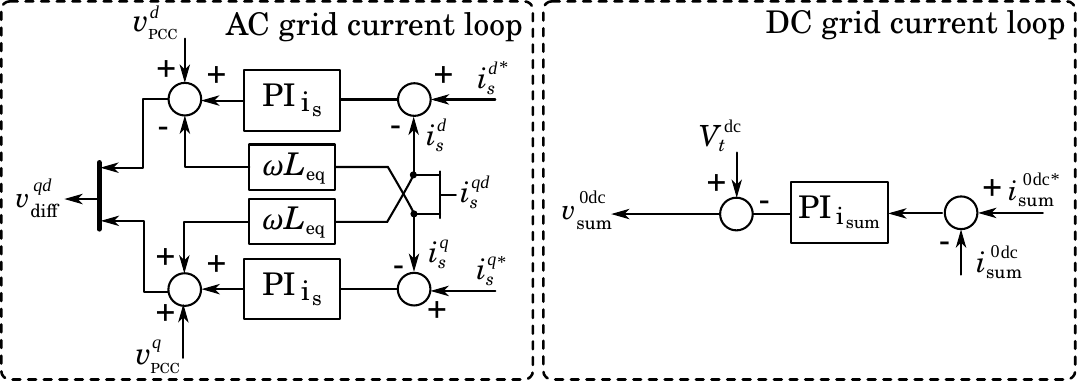}
	\caption{AC current control in $qd$-frame and DC current control. \label{fig:mmc_currcontrol}}
	\vspace{1em}
	\includegraphics[width=1\columnwidth]{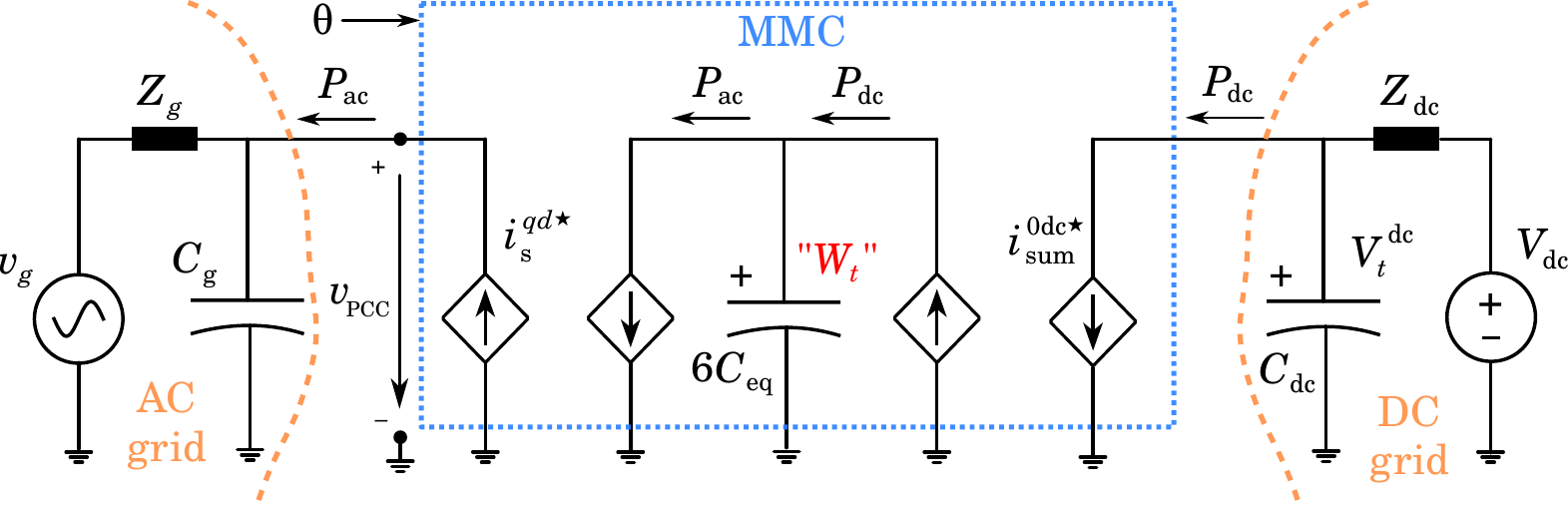}
	\caption{Macroscopic MMC model with DC current control and AC current control in a synchronous reference frame with angle $\theta$. The reference angle $\theta$ is given either by an outer AC-GFM control or a phase-locked loop (PLL). \label{fig:mmcabstractcurr}}
\end{figure}

We now assume that the current controllers in Fig.~\ref{fig:mmc_currcontrol} perfectly track their reference (i.e., $i^{qd}_s\!=\!i^{qd\star}_s$ and $i^{0\text{dc}}_{\text{sum}}\!=\!i^{0\text{dc}\star}_{\text{sum}}$) and that the internal energy balancing controllers ensure horizontal and vertical balancing. This allows neglecting the fast inner MMC dynamics and, assuming a lossless MMC, results in the model shown in Fig.~\ref{fig:mmcabstractcurr} and the total internal energy dynamics 
\begin{align}\label{eq:intdyn}
 \ddt W_t = P_{\text{dc}} - P_{\text{ac}},
\end{align}
with $P_{\text{ac}}= \tfrac{3}{2}{v^{qd}_{\text{diff}}}^\mathsf{T} i^{qd\star}_s$ and $P_{\text{dc}} = 3  v^{0\text{dc}}_{\text{sum}} i^{0\text{dc}\star}_{\text{sum}}$, i.e., we can control the total internal energy $W_t$ through the grid current reference on the DC side or AC side, or simultaneously through both sides (see Fig.~\ref{fig:watertank}).
\section{Review of single-port grid-forming control}\label{sec:sosgridform}
In practice a mixture of different controls is used to control the MMCs AC and DC terminals. The most common (single-port) GFM control structures form the voltage on one of the MMC terminals and can be broadly categorized into
\begin{itemize}
	\item DC-GFM that tracks a reference for $V^{\text{dc}}_t$ (e.g., $P_{\text{dc}}$-$V^\text{dc}_t$-droop) and controls the AC current to balance the internal energy $W_t$ (see Fig.~\ref{fig:detailed_control_dcforming} and Fig.~\ref{fig:mmcabstractdcform}),
\item AC-GFM that tracks a reference for $\theta$ and $V^\text{ac}_t$ (e.g., $P_\text{ac}$-$f$ and $Q$-$V^\text{ac}_t$ droop) and controls the DC current to balance the internal energy $W_t$ (see Fig.~\ref{fig:detailed_control_acforming} and Fig.~\ref{fig:mmcabstractacform}).
\end{itemize}
These controls aim to \emph{form} the grid voltage on one terminal of the MMC (i.e., either AC or DC) while balancing the internal energy $W_t$ by controlling the power flowing in and out of the other terminal of the MMC. In this setting, the terminal controlled for balancing the internal energy is assumed to be connected to a stiff grid (e.g., an infinite AC or DC bus). In the remainder of this section, we briefly review the prevalent implementations of AC-GFM control and DC-GFM control and discuss their limitations.

\subsection{DC-GFM voltage control and AC-GFL energy control}\label{subsec:DCform}
Starting from the MMC model in Fig.~\ref{fig:mmcabstractcurr} (i.e., assuming that the current controllers in Fig.~\ref{fig:mmc_internalbalcontrol} and Fig.~\ref{fig:mmc_currcontrol} perfectly track their reference), a standard DC-GFM droop control \cite{GSA+20} for MMCs can be implemented as follows. A phase-locked loop (PLL) is used to synchronize the MMC with the AC grid, i.e., to obtain phase angle $\theta$ and magnitude $V^{\text{ac}}_t$ of the AC grid-side voltage $v^{abc}_{\text{PCC}}$, and the total internal energy is controlled by a PI controller that provides an AC current reference to the underlying AC current control depicted in Fig.~\ref{fig:mmc_currcontrol}. A DC voltage PI control is used to regulate the DC voltage through the DC current (i.e., the DC current zero component $i_{\text{sum}}^{0{\text{dc}}\star}$) and $P_\text{dc}$-$V^\text{dc}_t$ droop is used to calculate the voltage reference.

\begin{figure}[!!tp]
	\centering
	\includegraphics[width=1\columnwidth]{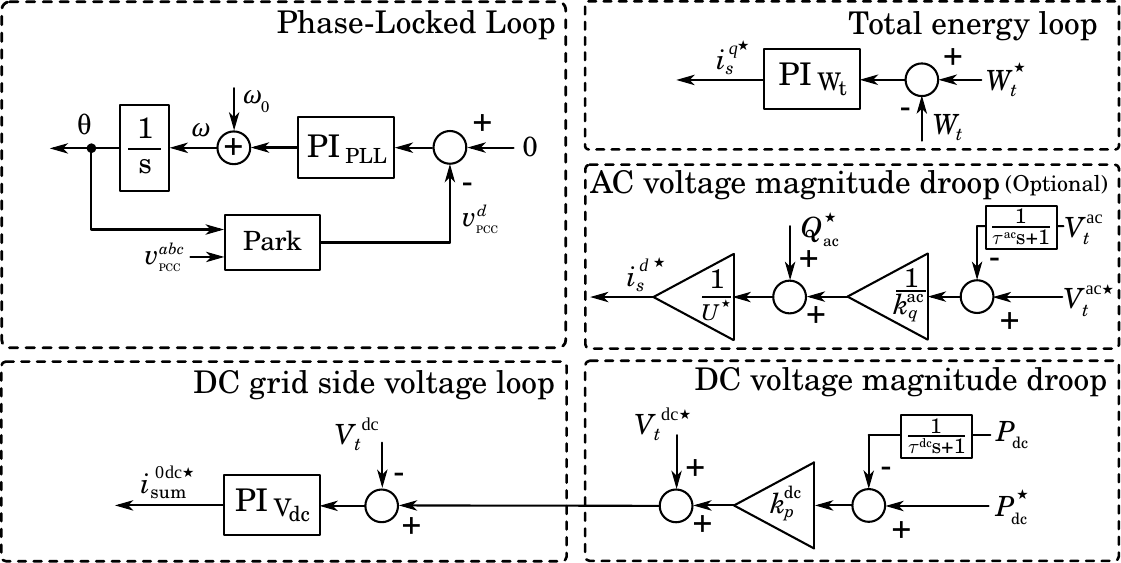}
	\caption{The standard DC-GFM/AC-GFL MMC control architecture forms the DC side voltage $V_t^\text{dc}$ through $P_{\text{dc}}$-$V_{\text{dc}}$ droop control and stabilizes the MMC internal energy $W_t$ through the AC current $i_s^q$.\label{fig:detailed_control_dcforming}}	
	\vspace{1em}
	\includegraphics[width=1\columnwidth]{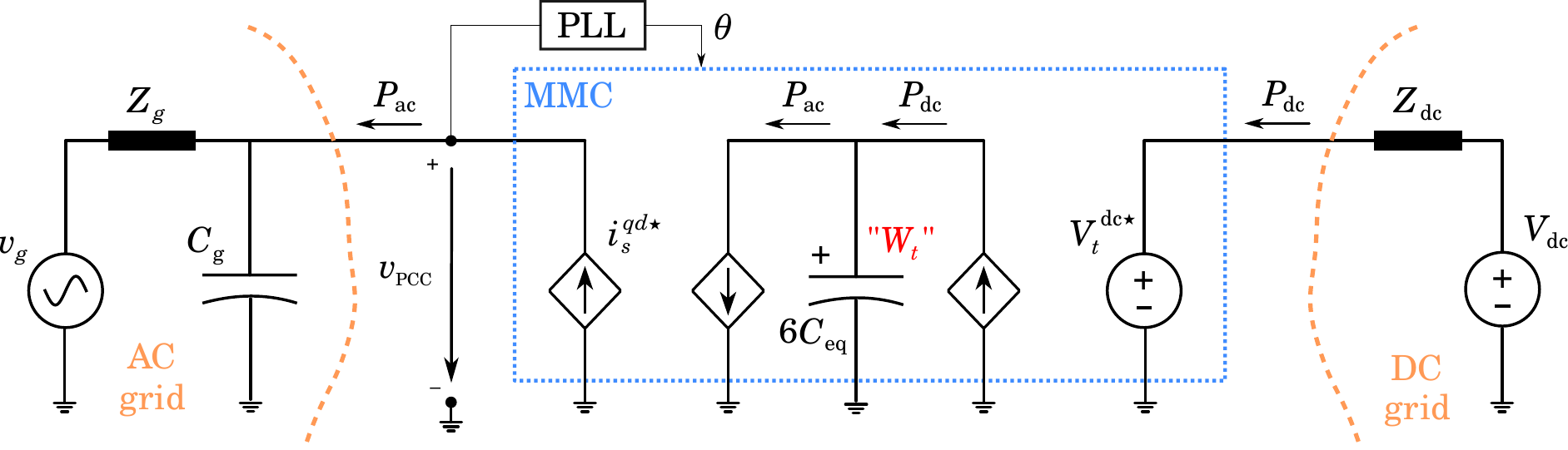}
	\caption{Macroscopic model of an DC-GFM/AC-GFL MMC. $V_t^{\text{dc}\star}$ and $i_s^{qd\star}$ are prescribed by the DC voltage magnitude droop control and total energy loop shown in Fig.~\ref{fig:detailed_control_dcforming}. \label{fig:mmcabstractdcform}}
\end{figure}

The overall DC-GFM/AC-GFL control is shown in Fig.~\ref{fig:detailed_control_dcforming} and the resulting macroscopic MMC model (i.e., only considering the MMCs total internal energy) is shown in Fig.~\ref{fig:mmcabstractdcform}. The macroscopic model in Fig.~\ref{fig:mmcabstractdcform} and controls in Fig.~\ref{fig:detailed_control_dcforming}  highlight that this approach forms a stable DC voltage (i.e., through $P_{\text{dc}}$-$V_{\text{dc}}$ droop control), but acts as a grid-following current source/load on the AC side that is controlled to stabilize the internal energy $W_t$ (i.e., assumes the AC grid to be stable).

Alternatively, the DC voltage PI control can be omitted and a measurement of $V^\text{dc}_t$ can be used to obtain the DC current reference $i_{\text{sum}}^{0{\text{dc}}\star}$ through $V^\text{dc}_t$-$P_\text{dc}$ droop. Notably, both variants cannot control the internal energy $W_t$ if the AC grid has no AC-GFM source.

\subsection{AC-GFM control and DC-GFL energy control}\label{subsec:ACform}
\begin{figure}[!!tp]
	\centering
	\includegraphics[width=1\columnwidth]{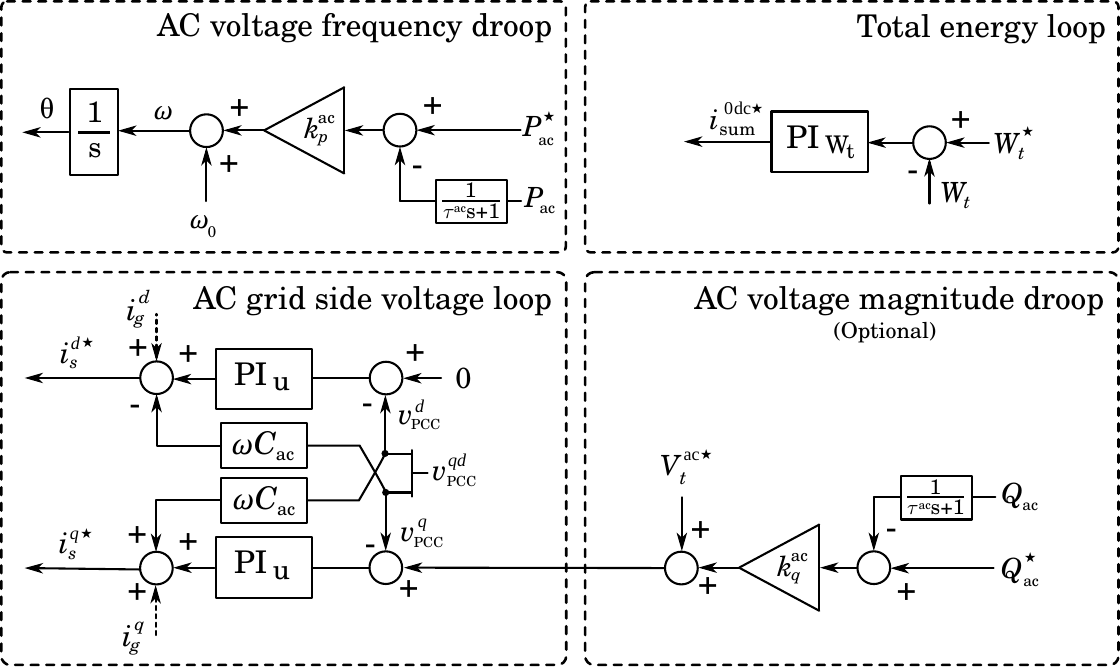}
	\caption{The standard AC-GFM/DC-GFL MMC control forms the AC voltage $v^\text{qd}_{\text{PCC}}$ using $P_{\text{ac}}$-$f$ droop control and stabilizes the MMC internal energy $W_t$ through the DC current $i_{\text{sum}}^{0{\text{dc}}}$. \label{fig:detailed_control_acforming}}
\vspace{1em}
	\includegraphics[width=1\columnwidth]{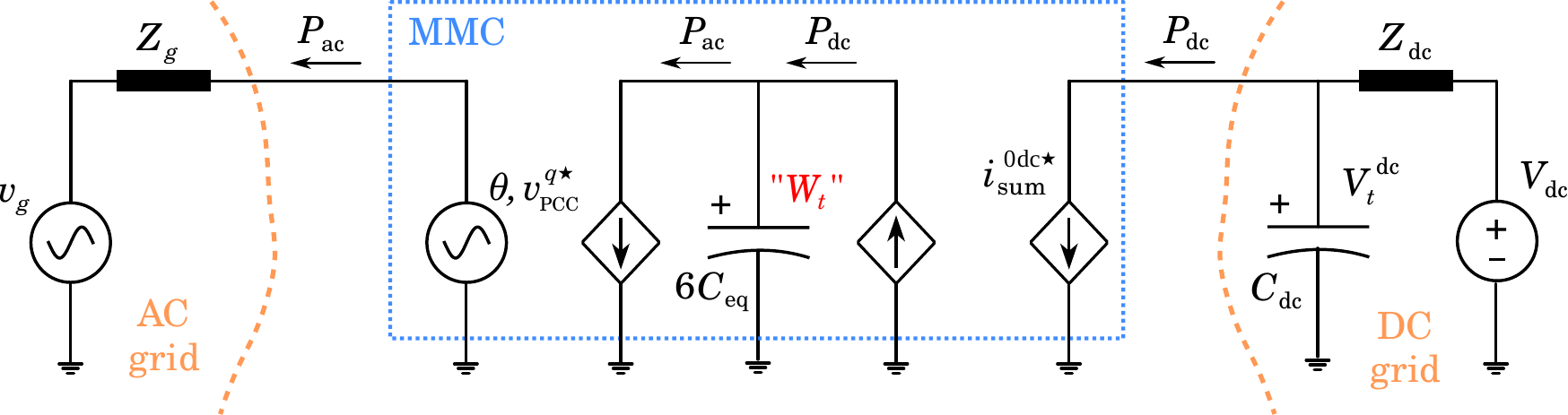}
	\caption{Macroscopic AC-GFM/DC-GFL MMC model. The total energy loop, $P_{\text{ac}}$-$f$ droop, and $Q_{\text{ac}}$-$V^{\text{ac}}_t$ droop shown in Fig.~\ref{fig:detailed_control_acforming} determine the references $i_{\text{sum}}^{0{\text{dc}\star}}$, $\theta$, and $v^{q\star}_{\text{PCC}}$.\label{fig:mmcabstractacform}}
\end{figure}

Building on the MMC model in Fig.~\ref{fig:mmcabstractcurr} (i.e., assuming that current controllers in Fig.~\ref{fig:mmc_internalbalcontrol} and Fig.~\ref{fig:mmc_currcontrol} perfectly track their reference), the standard AC-GFM control for MMCs is obtained by including a PI vector control that tracks a reference for the AC grid side voltage and uses the AC grid current injection ($i_s^{qd\star}$) as a control input. The voltage reference angle $\theta$ is determined using frequency droop control, i.e., based on the active power measurement, active power setpoint, and the $P_{\text{ac}}$-$f$ droop coefficient $k^{\text{ac}}_p$. In the remainder, we assume that volt-var droop is used to adjust the AC voltage magnitude reference $v^{q\star}_{\text{PCC}}$ as a function of the reactive power injection $Q_\text{ac}$. Finally, the total internal energy is controlled through the DC current ($i_{\text{sum}}^{0{\text{dc}}\star}$) using a PI controller.  The overall AC-GFM control is shown in Fig.~\ref{fig:detailed_control_acforming} and the resulting macroscopic MMC model (i.e., only considering the MMCs total internal energy) is shown in Fig.~\ref{fig:mmcabstractacform}. Note that the DC and AC side capacitances model the equivalent parasitic capacitance of the HVDC cable and HVAC transformer or line, rather than a large DC-link capacitor and AC filter capacitor as in two-level VSCs. Thus, the current $i_g$ (often used as a feed-forward signal in two-level VSCs) cannot be directly measured. Using the controls shown in Fig.~\ref{fig:detailed_control_acforming}, the MMC forms a stable AC voltage (i.e., through $P_{\text{ac}}$-$f$ droop control), but act as a grid-following current source/load on the DC side (see Fig.~\ref{fig:mmcabstractacform}) that is controlled to stabilize the internal energy $W_t$ under the assumption that the DC grid is stable irrespective of the DC power injection $P_{\text{dc}}$ of the MMC.

\subsection{Limitations of state-of-the art single-port GFM controls}\label{sec:sosgridform:disc}
State-of-the-art single-port DC-GFM and single-port AC-GFM approaches have significant limitations. To begin with, the IPC's AC and DC side controllers are implemented independently from each other (see AC and DC control sides in Fig.~\ref{fig:detailed_control_dcforming} and Fig.~\ref{fig:detailed_control_acforming}) and assume that a GFM resource always provides the required power. As a consequence, AC-GFM control may attempt to supply more power to an AC load than available on the DC side and collapse the MMCs internal energy while a control that coordinates the DC side, AC side, and energy control may be able to maintain stable operation (e.g., for frequency dependent loads). 

\begin{figure}[htbp]
	\centering
	\includegraphics[width=1\columnwidth]{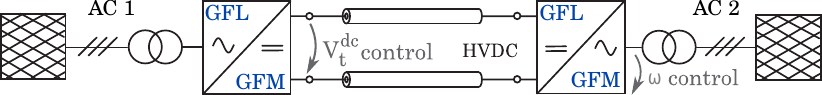}
	\caption{Standard GFL/GFM HVDC control architecture. \label{fig:HVDC}}
\end{figure}

More importantly, single-port GFM controls are specifically designed to leverage a stiff grid on the IPC's GFL terminal to stabilize the internal energy. Therefore, single-port GFM controls are not able to operate if the grid on the IPCs GFL terminal is not stiff due to, e.g., a loss of GFM resources or other contingencies. We emphasize that assigning GFM/GFL roles to the IPC terminals is non-trivial for complex grids-of-grids \cite{GSA+20} and, in future systems, significant challenges and complexities may arise even in relatively simple settings. For example, consider the point-to-point HVDC link in Fig.~\ref{fig:HVDC} that relies on stiff frequency control in system AC~1 to form the HVDC cable voltage which, in turn, is used to form the system AC~2. In this setup, contingencies in AC~1 or increased frequency volatility due to renewable integration can lead to a loss of PLL synchronization and challenge the conventional DC-GFM control structure by compromising its AC-based MMC energy control (see Fig.~\ref{fig:detailed_control_dcforming}). Analogously, losing the stiff DC grid implies losing the energy control and AC-GFM features of the AC-GFM IPC, as it is linked to a current controller that relies on a stiff DC system (see Fig.~\ref{fig:detailed_control_acforming}).

Overall, the availability of grid-forming resources and primary control reserves in different AC or DC subgrids may rapidly vary over time and cause single-port GFM controls with fixed GFM/GFL roles to fail. In this case, the standard control would have to be reconfigured within a few hundred milliseconds by, i.e., (i) identifying the new operating conditions, (ii) finding a suitable GFM/GFL role assignment, and (iii) coordinating a simultaneous mode transition of the IPC's controls.
To overcome these significant challenges and complexities, the next section proposes two dual-port GFM control structures that do not require GFM/GFL role assignments, form both DC and AC terminal voltages, and autonomously respond to changes in operating conditions and contingencies.
\section{Dual-port grid-forming control}\label{sec:dualgridform}
To overcome the limitations of single-port GFM control, this section presents two dual-port GFM controls that directly control the MMC's AC and DC terminal voltages and control the MMC's internal energy through AC-GFM and DC-GFM control. 

\begin{figure}[htbp]
	\centering
	\includegraphics[width=1\columnwidth]{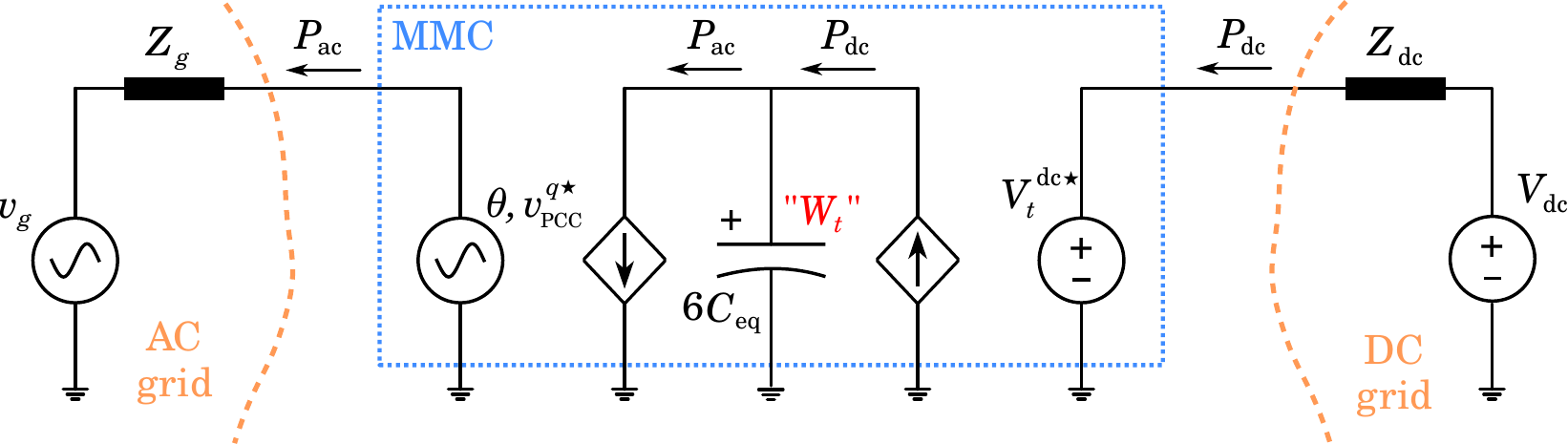}
	\caption{Macroscopic MMC model with dual-port voltage control. \label{fig:mmcabstractvolt}}
\end{figure}

To this end, the internal energy-balancing controls in Fig.~\ref{fig:mmc_internalbalcontrol} and inner current controls Fig.~\ref{fig:mmc_currcontrol} are combined with the DC grid side voltage control from Fig.~\ref{fig:detailed_control_dcforming} and AC grid side voltage control from Fig.~\ref{fig:detailed_control_acforming}. Assuming that the voltage controllers result in sufficiently fast and accurate voltage control\footnote{Similar timescale separation assumptions are required for standard DC-GFM and AC-GFM control (cf. \cite{SG2020}).} we obtain the MMC model in Fig.~\ref{fig:mmcabstractvolt} in which the AC voltage angle reference $\theta$, AC voltage magnitude reference $v^{q\star}_{\text{PCC}}$, DC voltage reference $V^{\text{dc}\star}_t$ remain as inputs for the dual-port GFM controls. For brevity of the presentation, we assume that volt-var droop is used to obtain the AC voltage magnitude reference $v^{q\star}_{\text{PCC}}$ (see Fig.~\ref{fig:detailed_control_acforming}).

\subsection{Hybrid power/energy droop control}
\begin{figure}[bp]
	\centering
	\includegraphics[width=1\columnwidth]{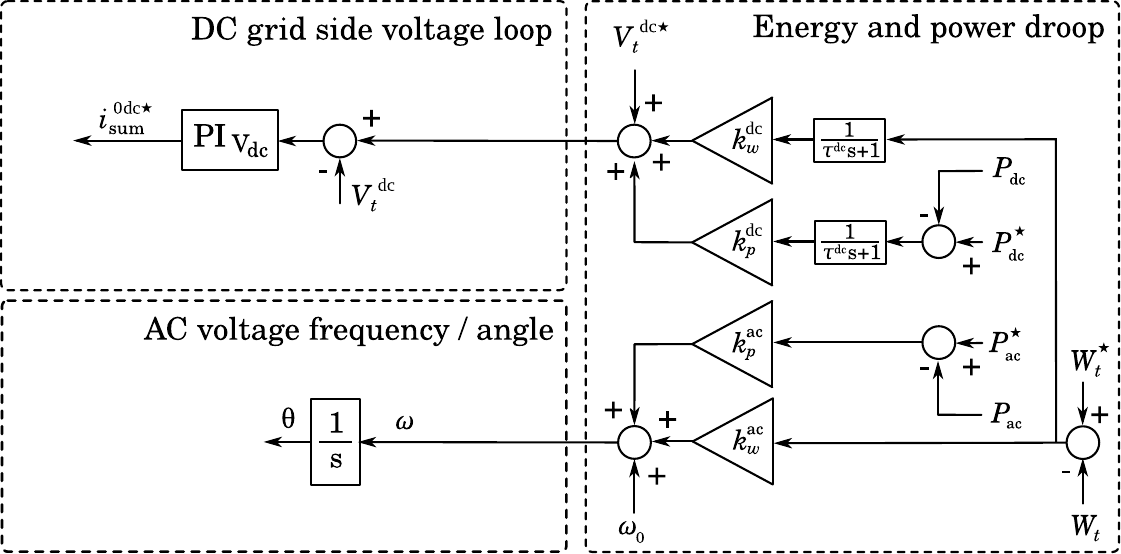}
	\caption{Hybrid power/energy droop control forms an AC and DC voltage that is adjusted based on droop control and a proportional energy control.\label{fig:detailed_control_hybriddr}}	
\end{figure}

We first consider the hybrid power/energy droop control
\begin{subequations}\label{eq:nonstiffform}
	\begin{align}
	\omega &= \omega^\star + k_{p}^{\text{ac}} (P^\star_{\text{ac}} - P_{\text{ac}})+k_{w}^{\text{ac}}(W_t - W^\star_t),\label{eq:anglet}\\
	V^{\text{dc}}_t&=\! V^{\text{dc}\star}_t + G_{p}^{\text{dc}}(s) (P_{\text{dc}}\!-\!P^\star_{\text{dc}})\!+\!G_{t}^{\text{dc}}(s) (W_t \!-\! W^\star_t),\label{eq:dcvoltt}
	\end{align} 
\end{subequations}
that combines DC and AC active power droop terms with an additional energy feedback inspired by \cite{CBB+2015,CGD2017,JAD18} (see Fig.~\ref{fig:detailed_control_hybriddr}). Where $G_{p}^{\text{dc}}(s) = \frac{k_{p}^{\text{dc}}}{\tau^{\text{dc}}s+1}$, and $G_{t}^{\text{dc}}(s) = \frac{k_{w}^{\text{dc}}}{\tau^{\text{dc}}s+1}$ are lowpass filters with time constant $\tau^\text{dc}$.
Here, $\omega^\star$, $V^{\text{dc}\star}_t$, $P^\star_{\text{ac}}$, $P^\star_{\text{dc}}$, and $W^\star_t$ denote the setpoints for frequency, DC voltage, AC active power, DC power, and the internal energy. Moreover, $k_{p}^{\text{ac}}$, $k_{p}^{\text{dc}}$, $k_{w}^{\text{ac}}$, and $k_{w}^{\text{dc}}$ are positive control gains resulting in $P_{\text{ac}}$-$f$, $P_{\text{dc}}$-$V^\text{dc}_t$, $W_t$-$f$, and $W_t$-$V^\text{dc}_t$ droop, respectively. Finally, $\tau^{\text{dc}}$ denotes a (small) low-pass filter time-time constant used to enforce time-scale separation between the outer dual-port GFM control and inner voltage and current controls.

Broadly speaking, this controller resembles AC and DC side droop control if $W_t \!-\! W^\star_t$ is small, i.e., adjusts AC frequency and DC voltage using active power measurements. However, the energy control terms will dominate the response if $W_t \!-\! W^\star_t$ is large and reduce (or increase) the AC frequency and the DC voltage to draw more (or inject less) power on the AC and DC side. For example, if a DC source stabilize $V^{\text{dc}}_t$, then $W_t$ is implicitly stabilized through the DC side (see Sec.~\ref{sec:stability}) and the frequency dynamics approximately resemble $P_{\text{ac}}$-$f$ droop control provided by AC-GFM/DC-GFL control. On the other hand, if the AC grid has stiff frequency regulation, then $W_t$ is implicitly stabilized through the AC side and the DC voltage dynamics approximately resemble $P_{\text{dc}}$-$V^\text{dc}_t$ droop control provided by AC-GFL/DC-GFM control.

\subsection{Energy-balancing control}
Conceptually, one may aim to form the AC and DC voltage while controlling the internal energy through the AC and DC side and suitable controls $G_w^{\text{ac}}(s)$ and $ G_w^{\text{dc}}(s)$, i.e., using
\begin{subequations}\label{eq:nopowersetpoint}
	\begin{align}
	\omega &= \omega^\star + G_w^{\text{ac}}(s) (W_t - W^\star_t),\label{eq:angletnps}\\
	V^{\text{dc}}_t&= V^{\text{dc}\star}_t + G_w^{\text{dc}}(s) (W_t - W^\star_t).\label{eq:dcvolttnps}
	\end{align} 
\end{subequations}

\begin{figure}[bp]
	\centering
	\includegraphics[width=1\columnwidth]{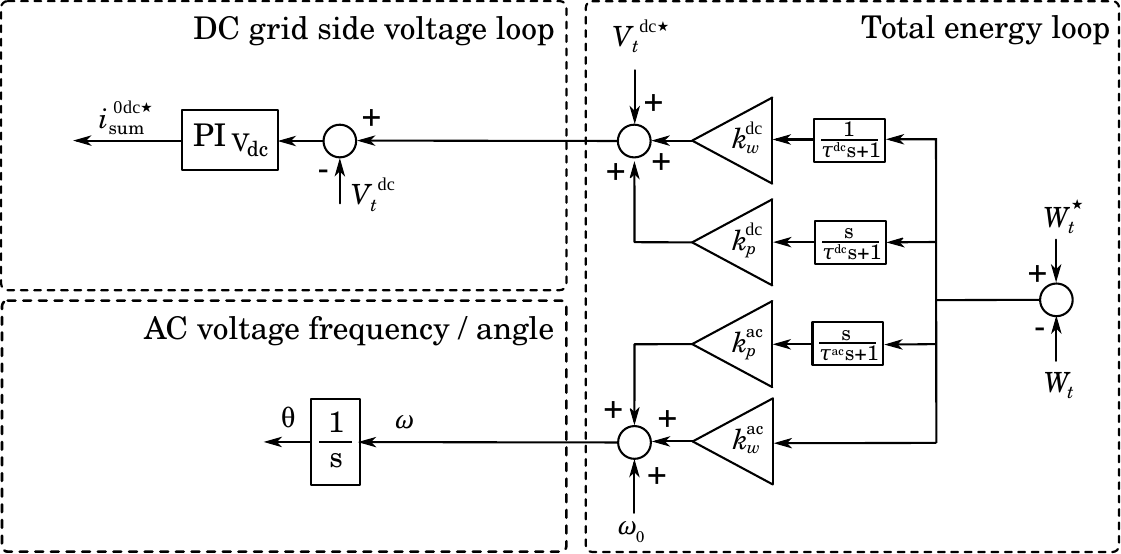}
	\caption{Energy-balancing control forms a AC and DC voltage that is adjusted through a realizable proportional-derivative energy control.\label{fig:detailed_control_enbal}}	
\end{figure}
For clarity of the presentation, we focus on the realizable proportional-derivative (PD) controls $G_w^{\text{ac}}(s) = k^{\text{ac}}_w + \tfrac{k_p^{\text{ac}} s}{\tau^{\text{ac}} s + 1}$ and $G_w^{\text{dc}}(s) = \tfrac{k_p^{\text{dc}} s + k^{\text{dc}}_w}{\tau^{\text{dc}} s + 1}$ with filter time constants $\tau^\text{ac}$ and $\tau^\text{dc}$  (see Fig.~\ref{fig:detailed_control_enbal}). However, we highlight that additional filters can be incorporated to meet control objectives beyond basic AC frequency and DC voltage stability (i.e., suppressing inter area oscillations). 

The (positive) control gains $k_{p}^{\text{ac}}$, $k_{p}^{\text{dc}}$, $k_{w}^{\text{ac}}$, $k_{w}^{\text{dc}}$  result in $\ddt W_t$-$f$, $\ddt W_t$-$V^\text{dc}_t$, $W_t$-$f$, and $W_t$-$V^\text{dc}_t$ droop, respectively. In other words, the $W_t$-$f$ and $W_t$-$V^\text{dc}_t$ droop stabilize the internal energy (e.g., lower or increase the AC frequency to lower or increase the AC power injection) and the $\ddt W_t$-$f$, $\ddt W_t$-$V^\text{dc}_t$ derivative-droop provides additional damping. While it may seem counter-intuitive to omit droop terms with power setpoints, we highlight that the objective is to form and stabilize the AC and DC grid voltages, not to transfer a fixed amount of power between the MMC's DC and AC terminals.

Nonetheless, energy-balancing control can be fully dispatched. The DC power injection $P_{\text{dc}}$ of an IPC at the nominal operating point is fully determined by $V^{\text{dc}\star}_t$ and in steady-state (i.e., $\ddt W_t=0$) it holds that $P^\star_{\text{dc}} = P^\star_{\text{ac}}$. In other words, the nominal steady-state power transfer through an IPC using \eqref{eq:nopowersetpoint} can be dispatched through $V^{\text{dc}\star}_t$ (see Sec.~\ref{sec:complextest}).

\subsection{Interpretation of dual-port grid-forming controls}
To provide an interpretation of \eqref{eq:nonstiffform} and \eqref{eq:nopowersetpoint}, we restrict our attention to $\tau^\text{ac}=\tau^\text{dc}=0$, i.e.,  no filtering of the DC voltage reference in \eqref{eq:nonstiffform} and ideal PD control in \eqref{eq:nopowersetpoint}.

Using  $P^{\text{ref}}_{\text{ac}}=P^\star_{\text{ac}}+\tfrac{k_{w}^{\text{ac}}}{k_{p}^{\text{ac}}} (W_t - W^\star_t)$ and $P^{\text{ref}}_{\text{dc}}= P^\star_{\text{dc}}+\tfrac{k_{w}^{\text{dc}}}{k_{p}^{\text{dc}}}(W_t - W^\star_t)$ we can write \eqref{eq:nonstiffform} in the form of standard droop control
\begin{subequations}\label{eq:droopequiv}
	\begin{align}
	\omega &= \omega^\star + k_{p}^{\text{ac}} (P^\text{ref}_{\text{ac}}- P_{\text{ac}}),\\ 	
	V^{\text{dc}}_t &=\! V^{\text{dc}\star}_t \!+\!  k_{p}^{\text{dc}}(P_{\text{dc}}\!-\!P^{\text{ref}}_{\text{dc}}).
	\end{align} 
\end{subequations}
In other words, the power setpoints $P^\star_{\text{ac}}$ and $P^\star_{\text{dc}}$ are modified to implement proportional control of $W_t$ through the AC and DC power references of the AC and DC droop controller.

Moreover, using $\tau^\text{ac}=\tau^\text{dc}=0$, we can rewrite \eqref{eq:nopowersetpoint} as
\begin{subequations}\label{eq:nopowersetpoint.olddiff}
	\begin{align}
		\omega &= \omega^\star + k_{p}^{\text{ac}} \ddt (W_t - W^\star_t) + k_{w}^{\text{ac}}(W_t - W^\star_t),\\
	V^{\text{dc}}_t &= V^{\text{dc}\star}_t +  k_{p}^{\text{dc}} \ddt (W_t \!-\! W^\star_t)\!+\!k_{w}^{\text{dc}}(W_t \!-\! W^\star_t).
	\end{align} 
\end{subequations}
Substituting $\ddt W_t= P_{\text{dc}}-P_{\text{ac}}$ from \eqref{eq:intdyn} into \eqref{eq:nopowersetpoint.olddiff} results in
\begin{subequations}\label{eq:nopowersetpoint.old}
	\begin{align}
		\omega &= \omega^\star + k_{p}^{\text{ac}} (P_{\text{dc}} - P_{\text{ac}}) + k_{w}^{\text{ac}}(W_t - W^\star_t),\label{eq:angletnps.old}\\
	V^{\text{dc}}_t &= V^{\text{dc}\star}_t +\!  k_{p}^{\text{dc}}(P_{\text{dc}}\!-\!P_{\text{ac}})\!+\!k_{w}^{\text{dc}}(W_t \!-\! W^\star_t),\label{eq:dcvolttnps.old}
	\end{align} 
\end{subequations}
i.e., for $\tau^\text{ac}=\tau^\text{dc}=0$, energy-balancing control \eqref{eq:nopowersetpoint} is approximated by hybrid droop control \eqref{eq:nonstiffform} with $P^\star_{\text{ac}}=P_{\text{dc}}$ and $P^\star_{\text{dc}}=P_{\text{ac}}$. 
However, despite their similarities, \eqref{eq:nopowersetpoint} and \eqref{eq:nopowersetpoint.old} result different dynamics during transients because \eqref{eq:intdyn} is an approximation for an ideal lossless MMC obtained by neglecting  fast internal dynamics.

\begin{remark}{\bf(Inertia response of dual-port GFM control)}\label{rem:inertia}
Conceptually, both hybrid energy/power droop control \eqref{eq:nonstiffform} and energy-balancing control \eqref{eq:nopowersetpoint} naturally provide a limited inertia response by linking the frequency at the converter AC terminal to the MMCs internal energy storage. In particular, if $k_p^\text{ac}=0$, the frequency solely depends on $W_t-W^\star_t$ just as the frequency in a synchronous machine is directly tied to its kinetic energy (see \cite{CGD2017} for a discussion in the context of two-level VSCs). However, the inertia is inherently limited by the MMC's internal energy storage (i.e., $30~\mathrm{ms}$ to $60~\mathrm{ms}$ at the rated power) and the fast power/energy damping induced by $k_p^\text{ac}>0$ potentially reduces the MMCs inertia response. 
\end{remark}

Finally, \eqref{eq:nopowersetpoint} and \eqref{eq:nopowersetpoint.old} result in a significantly different response from \eqref{eq:nonstiffform} during large contingencies. Using  \eqref{eq:nonstiffform}, the MMCs AC frequency and DC voltage respond linearly to (i) deviations  of the internal energy and (ii) deviations from the nominal power injection $P^\star_{\text{ac}}$ and $P^\star_{\text{dc}}$. However, whenever large contingencies occur, $P^\star_{\text{ac}}$, $P^\star_{\text{dc}}$, and $W^\star_t$ do not define a feasible operating point and 
a higher level operator has to identify the new situation and update the power setpoints in \eqref{eq:nonstiffform}. During this time (i.e., seconds to minutes), the power and energy droop terms in \eqref{eq:nonstiffform} represent conflicting objectives, and counter-intuitive IPC steady-state responses can be observed (see $t\geq 5.2~\mathrm{s}$ in Fig.~\ref{fig:MMCcomp_ps}).

In contrast, energy-balancing control \eqref{eq:nopowersetpoint} aims to balance the internal energy, AC frequency, and DC voltage irrespective of the deviation from the nominal power flow and simply rebalances the power flows (see $t\geq 5.2~\mathrm{s}$ in Fig.~\ref{fig:MMCcomp}).

\subsection{Steady-state response and control gains}\label{sec:ctrspecs}
The steady state relationship between AC frequency $\omega$, DC voltage $V^{\text{dc}\star}_t$, and internal energy $W_t$ of an IPC are characterized by the IPC controller equations with $\omega = \ddt \theta$ and/or $\ddt V^{\text{dc}}_{t}=0$. Using $P^\star_{\text{dc}} \!=\! P^\star_{\text{ac}}$ and substituting \eqref{eq:dcvoltt} into \eqref{eq:anglet} the total energy can be eliminated from the steady state map of an IPC with hybrid droop control \eqref{eq:nonstiffform} to obtain
\begin{align}\label{eq:nonstiffsteady}
\!\!\omega_k\!-\!\omega^\star_k \!=\!	\tfrac{k_{w}^{\text{ac}}}{k_{w}^{\text{dc}}} (V^{\text{dc}}_{t,m}\!-\! V^{\text{dc}\star}_{t,m})\!-\!(k_{p}^{\text{ac}}\!+\!\tfrac{k_{p}^{\text{dc}} k_{w}^{\text{ac}}}{k_{w}^{\text{dc}}})(P_{\text{ac}}\!-\!P^\star_{\text{ac}}).
 \end{align}
Next, letting $s=0$ in \eqref{eq:nopowersetpoint}, the steady state of a MMC with energy-balancing control \eqref{eq:nopowersetpoint} is given by
\begin{align}\label{eq:enbalsteady}
\omega_k-\omega^\star_k = k_{w}^{\text{ac}} (W_{t} - W^\star_{t}) = \tfrac{k_{w}^{\text{ac}}}{k_{w}^{\text{dc}}} (V^{\text{dc}}_{t,m}- V^{\text{dc}\star}_{t,m}).
\end{align}

In other words, using \eqref{eq:nopowersetpoint}, the steady-state relationship between the signals indicating  power imbalance in AC and DC networks (i.e., the AC frequency and DC voltage deviations) is straightforward and easily adjusted. We emphasize that \eqref{eq:nopowersetpoint} intentionally does not result in MMC steady-state droop behaviour. The IPC has no controllable power source and therefore cannot sustain steady-state droop behaviour. Instead, \eqref{eq:nopowersetpoint} maps power imbalances to all AC and DC sources that, if they provide droop, respond. In contrast, the mapping of steady-state AC frequency to DC voltage deviations using \eqref{eq:nonstiffform} also depends on the (post-contingency) power flow and can result in counter-intuitive responses to system imbalances in which the power and energy droop terms have opposite signs and cancel instead of resulting in the expected reduction (or increase) in AC frequency or DC voltage.

\begin{figure}[htpb!!]
	\centering
	\includegraphics[width=1\columnwidth]{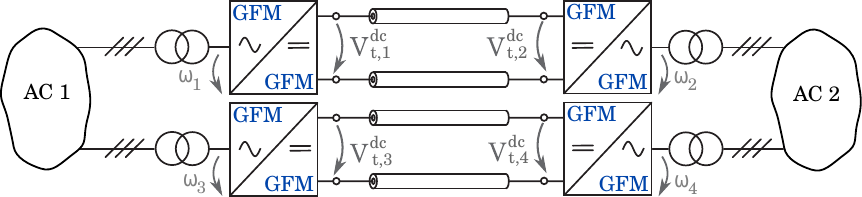}
	\caption{Two HVDC links interconnecting two AC networks. \label{fig:HVDCdual}}
\end{figure}

Based on the IPC steady-state equations, we can now discuss guidelines for selecting the control gains. We first consider the energy-balancing control \eqref{eq:nopowersetpoint} and note that the gains $k_w^{\text{ac}}$ and $k_w^{\text{dc}}$ can be chosen such that an expected worst-case frequency deviation maps to acceptable internal energy deviations and DC voltage deviations and vice-versa. Moreover, the gains  $k_p^{\text{ac}}$ and $k_p^{\text{dc}}$ and their associated time constants do not impact the steady-state and can be tuned to meet dynamic performance objectives. 
Next, we use the system in Fig.~\ref{fig:HVDCdual} as an example to illustrate the need to choose the control gains of different IPCs to ensure a consistent steady-state response. In particular, at steady state, $\omega_1=\omega_3$ and $\omega_2=\omega_4$ needs to hold. For brevity of the presentation we assume that $V^{\text{dc}\star}_{t,i}=V^{\text{dc}\star}_t$ for $i=\{1,\ldots,4\}$. Assuming negligible DC conductor losses, the voltage drop across the DC conductor is small, i.e., $V^{\text{dc}}_{t,1}-V^{\text{dc}\star}_{t}$ and $V^{\text{dc}}_{t,2}-V^{\text{dc}\star}_{t}$ are approximately identical. Under this assumption,\eqref{eq:enbalsteady} results in $\alpha_1(\omega_1 - \omega^\star_1) = \alpha_2(\omega_2 - \omega^\star_2)$ and $\alpha_3(\omega_3 - \omega^\star_3) = \alpha_4(\omega_4 - \omega^\star_4)$, where $\alpha_l\coloneqq k_{w,l}^{\text{ac}}/k_{w,l}^{\text{dc}}$. It follows that $\omega_1=\omega_3$ and $\omega_2=\omega_4$ holds for the system in Fig.~\ref{fig:HVDCdual} if $\alpha_1 / \alpha_2=\alpha_3 / \alpha_4$, $\omega^\star_1=\omega^\star_3$, and $\omega^\star_2=\omega^\star_4$. More generally, along any path connecting two AC networks, the products of the gain $\alpha_k / \alpha_l$ associated with any AC-DC-AC conversion need to be identical and each converter connected to the same AC network needs to use the same frequency set-point. In contrast, when using \eqref{eq:nonstiffform}, the dependence on power deviations (cf. \eqref{eq:nonstiffsteady}) and, thereby, on the IPC and generator power setpoints and droop coefficients, precludes a compact analysis. Instead, for the purpose of this study, we recommend to select the gains $k_p^{\text{ac}}$ (or $k_p^{\text{dc}}$) to ensure a consistent droop response when internal energy deviations are small, i.e., to be approximately the same for each AC (or DC) network and compatible with the prevailing droop setting of AC and DC generators. Moreover, we select $k_w^{\text{ac}}$ (or $k_w^{\text{dc}}$) such that the expected worst-case frequency and power (or DC voltage and power) deviation maps to acceptable internal energy deviations and the ratios $\alpha_k/\alpha_l$ associated any AC-DC-AC conversion are consistent between AC networks.

\section{Small-signal stability analysis}\label{sec:stability}
\begin{figure}[htpb!!]
	\centering
	\includegraphics[width=0.9\columnwidth]{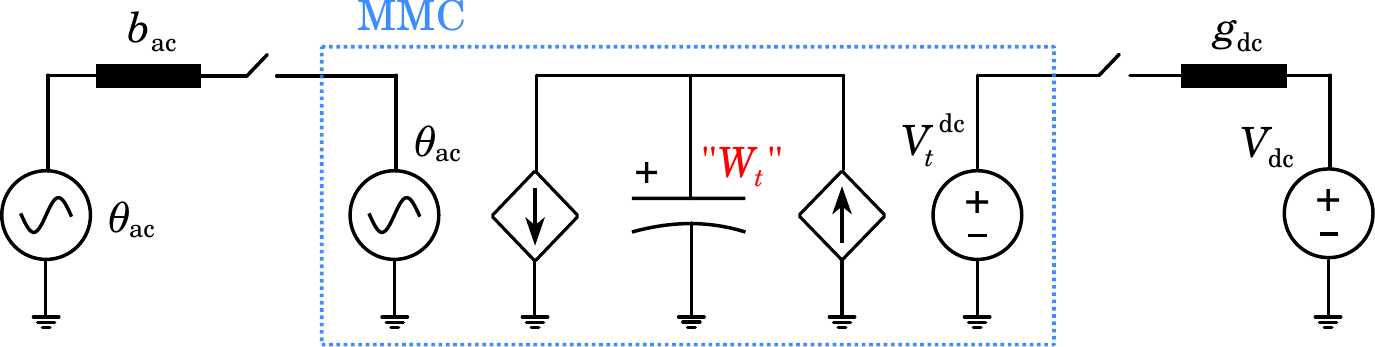}
	\caption{System used for the simplified stability analysis. \label{fig:simplestab}}
\end{figure}

Before proceeding to validate and illustrate the features of dual-port GFM control in simulation studies, we consider the simplified setup of an MMC interconnecting a DC grid and AC grid (see Fig.~\ref{fig:simplestab}) to analyze small-signal stability of the controllers \eqref{eq:nonstiffform} and \eqref{eq:nopowersetpoint} dual-port GFM control.
For the purpose of this analysis, we assume that the filter time constant in \eqref{eq:dcvoltt} and \eqref{eq:dcvolttnps} is negligible (i.e., $\tau^\text{dc}=0$) and that MMCs internal dynamics are fast and can be neglected. We linearize the system dynamics at the zero power flow solution to obtain% the small-signal model
\begin{align}\label{eq:acsidemod}
 P_{\text{ac}}=b_{\text{ac}}(\theta-\theta_{\text{ac}}), \quad
  \ddt \theta_{\text{ac}} = \omega^\star + k_{\text{ac}} P_{\text{ac}},
\end{align}
where $b_{\text{ac}} \in \mathbb{R}_{>0}$ is the AC network susceptance, $\theta_{\text{ac}}$ is the angle of the AC source and $k_{\text{ac}}  \in \mathbb{R}_{\geq 0}$ is its droop coefficient. Moreover, $P^\star$ is the MMC power setpoint, and $P_{\text{ac}}$ is the power flowing out of the MMC and into the AC grid. Similarly, the DC grid is modeled by an equivalent network and voltage source and linearizing the DC power flow equation results in
\begin{align}\label{eq:dcsidemod}
P_{\text{dc}}=g_{\text{dc}}(V_{\text{dc}}-V^{\text{dc}}_t), \quad 
%  V_{\text{dc}} = V^{\text{dc}\star}_{t}  + k_{\text{dc}} (P_{\text{dc}}-P^\star),
 V_{\text{dc}} = V^{\text{dc}\star}_{t}  + k_{\text{dc}} P_{\text{dc}},
\end{align} 
where $g_{\text{dc}} \in \mathbb{R}_{>0}$ is the DC network conductance, 
$V_{\text{dc}}$ is the dc source voltage and $k_{\text{dc}} \in \mathbb{R}_{\geq 0}$ is its droop coefficient. 

Our analytical stability result is summarized in Theorem \ref{thm:mmcstab}.
\begin{theorem}{\bf(MMC Stability)}\label{thm:mmcstab}
 Consider the MMC inernal energy dynamics \eqref{eq:intdyn} in closed loop with either (i) the  control \eqref{eq:nonstiffform} with $k_{w}^{\text{ac}}>k_{p}^{\text{ac}}$, or (ii) the control \eqref{eq:nopowersetpoint}. Moreover, assume that $\tau^\text{ac}=\tau^\text{dc}=0$ and $P^\star_{\text{ac}} = P^\star_{\text{dc}}=0$. The internal energy $W_t$, AC frequency $\omega$, and DC voltage $V^{\text{dc}}_t$ are asymptotically stable if the MMC is connected (i) only to the AC network \eqref{eq:acsidemod}, (ii) only to the DC network \eqref{eq:dcsidemod}, or (iii) to both networks.
\end{theorem}
\begin{IEEEproof}
We begin by deriving the closed-loop dynamics for \eqref{eq:intdyn}, the controller \eqref{eq:nonstiffform} with $\tau^\text{dc}=0$, \eqref{eq:acsidemod}, and \eqref{eq:dcsidemod}. Substituting $V_{\text{dc}}-V^{\text{dc}}_t$ from \eqref{eq:dcvoltt} into $P_{\text{dc}}$ in \eqref{eq:dcsidemod} and solving for $P_{\text{dc}}$ results in  $P_{\text{dc}} = -\kappa_{\text{dc}} \Delta W_{t}$, where $\Delta W_{t}=W_t - W^\star_t$ and $\kappa_{\text{dc}}=\frac{k_{w}^{\text{dc}}}{1+g_{\text{dc}}k_{\text{dc}}+g_{\text{dc}}k_{p}^{\text{dc}}}$. Next, using $\delta = \theta-\theta_{\text{ac}}$, $P_\text{ac}=b_{\text{ac}} \delta$, \eqref{eq:acsidemod}, \eqref{eq:anglet}, \eqref{eq:intdyn}, and $P_{\text{dc}} = -\kappa_{\text{dc}} \Delta W_{t}$, we obtain 
\begin{subequations}\label{eq:nonstiffclosedloop}
 \begin{align}
 \ddt \delta &= -(k_{p}^{\text{ac}}+k_{\text{ac}})b_{\text{ac}} \delta + k_{w}^{\text{ac}} \Delta W_t,\\
 \ddt \Delta W_t &= - \kappa_{\text{dc}} \Delta W_t -b_{\text{ac}} \delta.
\end{align}
\end{subequations}
Connecting only the AC network \eqref{eq:acsidemod} results in \eqref{eq:nonstiffclosedloop} with $\kappa_{\text{dc}}=0$. If only the DC network \eqref{eq:dcsidemod} is connected the dynamics reduce to $\ddt \Delta W_t = - \kappa_{\text{dc}} \Delta W_t$. 

By the Routh-Hurwitz criterion, all three cases are stable for all positive control gains. Next, we consider the controller \eqref{eq:nopowersetpoint} with $\tau^\text{ac} = \tau^\text{dc}=0$ in closed loop with \eqref{eq:intdyn}, \eqref{eq:acsidemod}, and \eqref{eq:dcsidemod}. Substituting $V_{\text{dc}}-V^{\text{dc}}_t$ from \eqref{eq:dcvolttnps} into $P_{\text{dc}}$ in \eqref{eq:dcsidemod}, and solving for $P_{\text{dc}}$ results in  $P_{\text{dc}} = -\gamma_{\text{dc}} k_{p}^{\text{dc}} \ddt \Delta W_{t}-\gamma_{\text{dc}} k_{w}^{\text{dc}} \Delta W_t$, where $\gamma_{\text{dc}}=g_{\text{dc}} (1-\frac{k_{\text{dc}} g_{\text{dc}}}{k_{\text{dc}} g_{\text{dc}}+1}) >0$. Moreover, \eqref{eq:angletnps} can be rewritten as 
$\ddt (\theta - k_{p}^{\text{ac}} \Delta W_t) =\omega^\star  + k_{w}^{\text{ac}} \Delta W_t$ and, letting $\vartheta = \theta - k_{p}^{\text{ac}} \Delta W_t - \theta_{\text{ac}}$, the resulting dynamics are given by
\begin{subequations}\label{eq:nopowersetpointclosedloop}
 \begin{align}
  \ddt \vartheta&= -k_{\text{ac}} b_{\text{ac}} \vartheta + (k_{p}^{\text{ac}} - k_{w}^{\text{ac}}) \Delta W_t,\\
  \ddt \Delta W_t &= - \tfrac{\gamma_{\text{dc}} k_{w}^{\text{dc}}+b_{\text{ac}} k_{w}^{\text{ac}}}{1+\gamma_{\text{dc}} k_{p}^{\text{dc}}}\Delta W_t + \tfrac{b_{\text{ac}}}{1+\gamma_{\text{dc}} k_{p}^{\text{dc}}} k_{p}^{\text{dc}}\vartheta.
 \end{align}
\end{subequations}
Connecting only the AC network \eqref{eq:acsidemod} results in \eqref{eq:nopowersetpointclosedloop} with $\gamma_{\text{dc}}=0$. If only the DC network \eqref{eq:dcsidemod} is connected, the dynamics reduce to $\ddt \Delta W_t = - \tfrac{\gamma_{\text{dc}} k_{w}^{\text{dc}}}{1+\gamma_{\text{dc}} k_{p}^{\text{dc}}}\Delta W_t$. By the Routh-Hurwitz criterion, all three cases are asymptotically stable if $k_{w}^{\text{ac}}>k_{p}^{\text{ac}}$. The Theorem directly follows from the fact that the MMC AC frequency $\omega$ and DC voltage $V^{\text{dc}}_t$ are asymptotically stable if $W_t$ is asymptotically stable.
\end{IEEEproof}

The proof uses the fact that the dual-port GFM control \eqref{eq:nonstiffform} can be interpreted as proportional energy control on the DC terminal and proportional energy control through a low-pass filter on the AC terminal (see \eqref{eq:nonstiffclosedloop}). Moreover, the dual-port GFM control \eqref{eq:nopowersetpoint} can be interpreted as proportional derivative energy control through the MMC's DC terminal and through a low pass filter on the AC terminal (see \eqref{eq:nopowersetpointclosedloop}). 

We emphasize that the system in Fig.~\ref{fig:simplestab} with standard DC-GFM control (cf. Sec.~\ref{subsec:DCform}) is only stable if the AC network is stabilized by the AC source $U_1$. With standard AC-GFM control (cf. Sec.~\ref{subsec:ACform}) it is only stable if the DC network is stabilized by the DC source $V_4$. In contrast, Theorem \ref{thm:mmcstab} establishes that the system in Fig.~\ref{fig:simplestab} using the dual-port GFM control \eqref{eq:nonstiffform} or \eqref{eq:nopowersetpoint} is stable if either the DC or AC grid are stable. This theoretical result will be verified in Sec.~\ref{sec:case_study:simp} using a detailed MMC and network model.

\section{Case Studies \& Comparisons}\label{sec:case_study}
\subsection{MMC and network model}\label{sec:case_study:model}
In the following case studies we use an MMC model with $400$ submodules per arm, ideal switches, and nearest level modulation with reduced switching-frequency (see e.g., \cite{QZL11}). We model DC cables using three parallel $\pi$ branches to accurately represents the frequency-dependent behavior of DC cables \cite{BDS16} and use the standard $\pi$-model to model AC transmission lines. The parameters are summarized in Table \ref{tab:case_params}.
\begin{table}[!!!pb]
\centering
\caption{Case study parameters \label{tab:case_params}}
\begin{tabular}[c]{lccccccl}
\hline\multicolumn{6}{c}{\cellcolor{gray!20}\textbf{MMC parameters} \cite{Prieto-Araujo2017a}}\\ \hline
\multicolumn{3}{c}{\textbf{Parameter}}                  & \textbf{Symbol}    & \textbf{Value} & \textbf{Units}	\\ \hline
\multicolumn{3}{c}{Rated (base) AC-side voltage}                     & $U_N$              & 320            & kV 	\\
\multicolumn{3}{c}{Rated (base) DC-side voltage}                   & $V^{\textrm{dc}}_N$           & $\pm$320       & kV  			\\
\multicolumn{3}{c}{Transformer leakage impedance} & $R_T$+j$L_T$ & $0.004$+j$0.15$ & pu \\ 
\multicolumn{3}{c}{Phase reactor impedance}              	& $R_s$+j$L_s$       & $0.005$+j$0.1$      & pu              \\
\multicolumn{3}{c}{Arm reactor impedance}               & $R_a$+j$L_a$       & $0.01$+j$0.2$     & pu              \\
\multicolumn{3}{c}{Converter modules per arm}           & $N_{\textrm{arm}}$          & 400            & -               \\
\multicolumn{3}{c}{Average module voltage}              & $V_{\textrm{SM}}$           & 1.6            & kV              \\
\multicolumn{3}{c}{Submodule capacitance}              & $C_{\textrm{SM}}$           & 8             & mF              \\
\hline\multicolumn{6}{c}{\cellcolor{gray!20}\textbf{DC cable parameters} \cite{F17}}\\ \hline
\textbf{Symbol} & \textbf{Value} 			& \textbf{Units} 	& \textbf{Symbol}	& \textbf{Value} 			& \textbf{Units}  \\ \hline
$r_1$        	& $0.1265$     & $\Omega$/km		& $l_1$        		& $0.2644$		& mH/km	\\
$r_2$        	& $0.1504$     & $\Omega$/km   	& $l_2$	       		& $7.2865$     & mH/km	\\
$r_3$        	& $0.0178$     & $\Omega$/km   	& $l_3$          	& $3.6198$     & mH/km	\\
$c$  	      	& $0.1616$     & $\mu$F/km     			& $g$     			& $0.1015$    & $\mu$S/km	\\
\hline\multicolumn{6}{c}{\cellcolor{gray!20}\textbf{AC line parameters}}\\ \hline
\textbf{Symbol} & \textbf{Value} 			& \textbf{Units} 	& \textbf{Symbol}	& \textbf{Value} 			& \textbf{Units}  \\ \hline
$r_\ell$       	& $0.08$     & $\Omega$/km		& $l_\ell$        		& $0.8$		& mH/km	\\
$c$  	      	& $0.012$     & $\mu$F/km     			&      			&     & 	\\
\end{tabular}
\end{table}

\subsection{Case study: test system with single IPC}\label{sec:case_study:simp}
\begin{figure}[t!!]
    \centering
	\includegraphics[width=0.9\columnwidth]{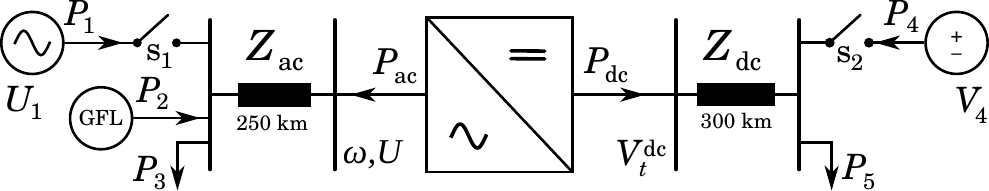}
	\vspace{0.1em}
	\caption{Test system with a single IPC, grid-forming AC source, grid-following AC source, and grid-forming DC source. \label{fig:simptest}}
\end{figure}

\begin{figure*}[h!!]
\includegraphics[width=1\textwidth]{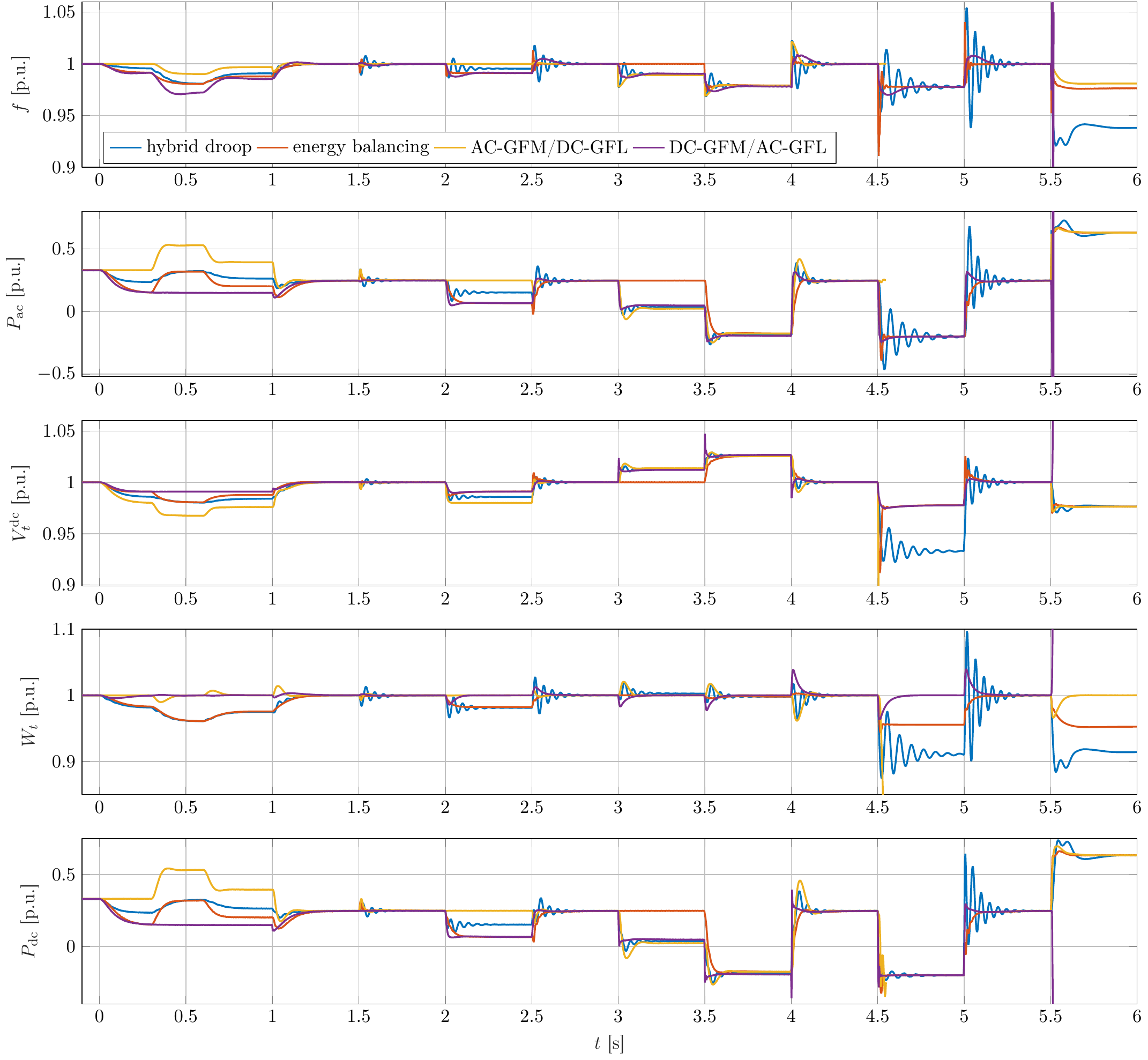}
	\caption{Response of different GFM controls to load/generation changes, setpoint updates, and contingencies in the system shown in Fig.~\ref{fig:simptest}. \label{fig:MMCsimp}}
\end{figure*}
To illustrate the main features of the proposed dual-port GFM controls we first consider the test system depicted in Fig.~\ref{fig:simptest} that consists of a $500~\mathrm{MW}$ MMC interconnecting a $320~\mathrm{kV}$ / $50~\mathrm{Hz}$ AC network and a $640~\mathrm{kV}$ DC network representing a common $\pm 320~\mathrm{kV}$ symmetrical monopole. At the rated power the MMC submodules provide approximately $49.15~\mathrm{ms}$ of electrostatic energy storage. The AC network contains a grid-forming source (i.e., AC voltage source) with $P_\text{ac}$-$f$ droop, an AC grid-following power source (current source with PLL), and a constant power load. 
\begin{figure*}[h!!]
\includegraphics[width=1\textwidth]{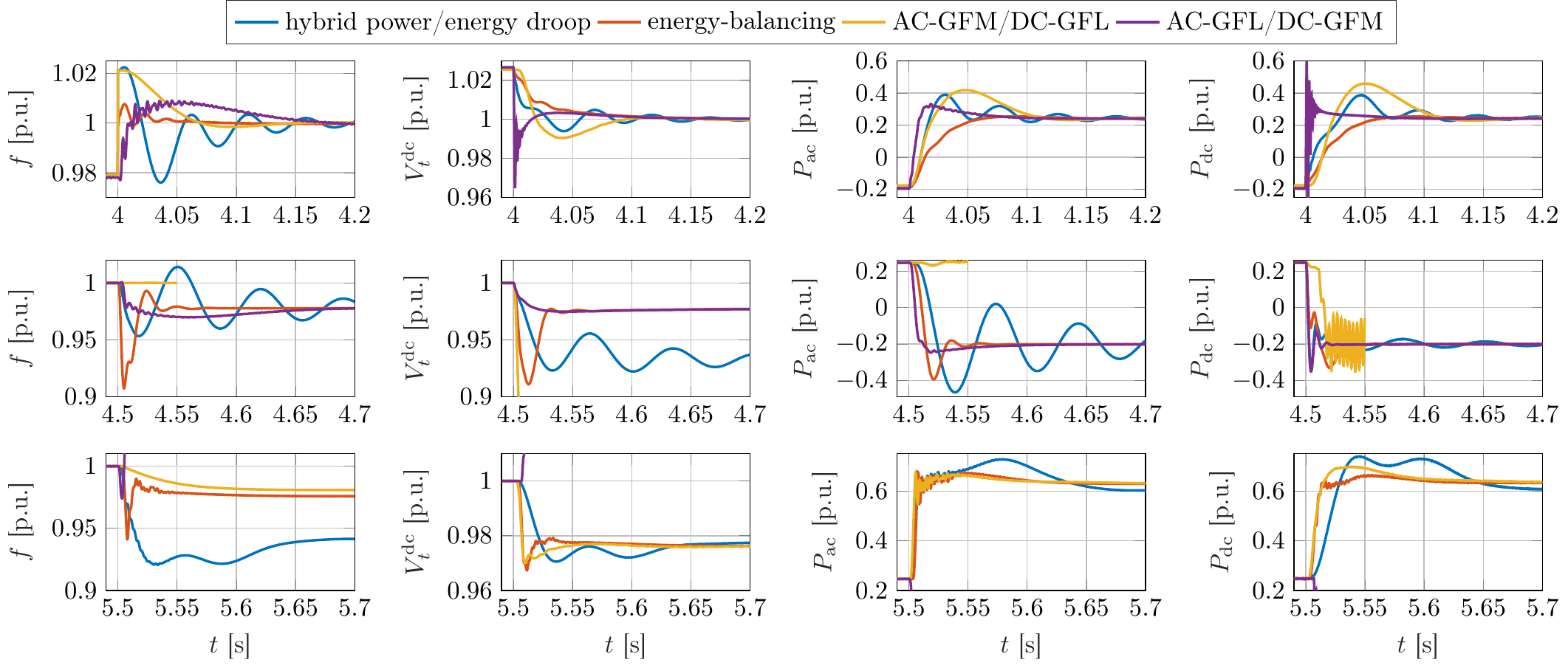}
	\caption{{Detailed view of the redispatch (first row), disconnection of the DC source $V_4$ (second row), and disconnection the AC source $U_1$ shown in Fig.~\ref{fig:MMCsimp}. Notably, after a brief transient (approx. one cycle) energy-balancing control converges to the response of the appropriate single-port GFM control when $V_4$ and $U_1$ are disconnected.}\label{fig:MMCsimp_zoom}}
\end{figure*}
The DC network contains a grid-forming power source (i.e., DC voltage source) with $P_{\text{dc}}$-$V_{\text{dc}}$ droop and a constant power load. The two GFM sources can be disconnected to investigate the DC and AC-GFM capabilities in different system configurations and the grid-following AC power source is used to verify that MMC can form an AC voltage that is sufficiently stable for renewable AC grid-following resources to operate. Simulation results for the two proposed control strategies as well as standard AC-GFM and DC-GFM control are shown in Fig.~\ref{fig:MMCsimp}. {Moreover, Fig.~\ref{fig:MMCsimp_zoom} provides a more detailed view of three events.}

All controls are combined with volt-var droop and the control parameters were selected based on the discussion in Sec.~\ref{sec:ctrspecs} and can be found in Table~\ref{table:ctrpar}.  The changes to setpoints, loads, and the system topology throughout the simulation are summarized in Table~\ref{table:simptest}. Initially, load and generation are balanced and the system operates at the nominal AC frequency / voltage and DC voltage. 
\begin{table}[b!]
\begin{center}
\caption{GFM control parameters: single IPC test system \label{table:ctrpar}}\vspace{0.1em}
\addtolength{\tabcolsep}{-3.5pt}    
 \begin{tabular}{l| c c c c c c c}
 Control & $k^\text{ac}_p$\,[pu] & $k^\text{ac}_q$\,[pu] & $k^\text{dc}_p$\,[pu] & $k^\text{ac}_w$\,[pu] & $k^\text{dc}_w$\,[pu] & $\tau^\text{ac}$\,[s] & $\tau^\text{dc}$\,[s]\\ [0.5ex] 
\hline
DC-forming & \textendash &  $0.05$ & $0.05$ & \textendash & \textendash & $1$ & $0.01$ \\
AC-forming & $0.05$ &  $0.05$ & \textendash & \textendash & \textendash & $0.04$ & \textendash \\
hybrid droop & $0.05$ &  $0.05$ & $0.05$ & $0.5$ & $0.5$ & \textendash & $0.01$ \\
energy-balan. & $0.0125$ &  $0.05$ & $0.025$ & $0.5$ & $0.5$ & $0.001$ & $0.01$ \\
\end{tabular}
\vspace{0.5em}
\caption{Dispatch and load values in per-unit (rounded)\label{table:simptest}}
 \begin{tabular}{l| c c c c c c c c c c}
 $t$ [s] & $P^\star_1$ & $U_1$ & $P_2$ & $P_3$ & $P^\star_4$  & $V_4$ & $P_5$ & $P^\star_{\text{MMC}}$ & $\omega^\star$ & $V^{\text{dc}\star}_t$ \\ [0.5ex] 
\hline
init. & $0.15$ &  $1$ & $0.35$ & $-0.82$ & $0.5$ & $1$ & $-0.1$ & $0.33$ & $1$ & $1$\\
$0$ & $0.15$ &  $1$ & $0.35$ & $-0.82$ & $0.5$ & $1$ & $\mathbf{-0.5}$ & $0.33$ & $1$ & $1$\\
$0.3$ & $0.15$ &  $1$ & $0.35$ & $\mathbf{-1.2}$ & $0.5$ & $1$ & $-0.5$ & $0.33$ & $1$ & $1$\\
$0.6$ & $0.15$ &  $1$ & $0.35$ & $\mathbf{-0.94}$ & $0.5$ & $1$ & $\mathbf{-0.2}$ & $0.33$ & $1$ & $1$\\ 
$1$ & $\mathbf{0.35}$ &  $1$ & $0.35$ & $-0.94$ & $0.5$ & $1$ & $-0.2$ & $\mathbf{0.25}$ & $1$ & $1$\\  
$1.5$ & $0.35$ &  $\mathbf{0.9}$ & $0.35$ & $-0.94$ & $0.5$ & $1$ & $-0.2$ & $0.25$ & $1$ & $1$\\  
$2$ & $0.35$ &  $0.9$ & $0.35$ & $-0.94$ & $0.5$ & $\mathbf{0.98}$ & $-0.2$ & $0.25$ & $1$ & $1$\\  
$2.5$ & $0.35$ &  $\mathbf{1}$ & $0.35$ & $-0.94$ & $0.5$ & $\mathbf{1}$ & $-0.2$ & $0.25$ & $1$ & $1$\\
$3$ & $0.35$ &  $1$ & $0.35$ & $-0.94$ & $0.5$ & $1$ & $-0.2$ & $\mathbf{-0.2}$ & $1$ & $1$\\
$3.5$ & $0.35$ &  $1$ & $0.35$ & $-0.94$ & $0.5$ & $1$ & $-0.2$ & $-0.2$ & $\mathbf{0.98}$ & $\mathbf{1.03}$\\
$4$ & $0.35$ &  $1$ & $0.35$ & $-0.94$ & $0.5$ & $1$ & $-0.2$ & $\mathbf{0.25}$ & $\mathbf{1}$ & $\mathbf{1}$\\
$4.5$ & $0.35$ &  $1$ & $0.35$ & $-0.94$ & \textbf{dis.} & \textbf{dis.} & $-0.2$ & $0.25$ & $1$ & $1$\\
$5$ & $0.35$ &  $1$ & $0.35$ & $-0.94$ & $\mathbf{0.5}$ & $\mathbf{1}$ & $-0.2$ & $0.25$ & $1$ & $1$\\
$5.5$ & \textbf{dis.} &  \textbf{dis.} & $0.35$ & $-0.94$ & $0.5$ & $1$ & $-0.2$ & $0.25$ & $1$ & $1$
\end{tabular}
\end{center}
\end{table}

At $t=0~\mathrm{s}$, the DC load is increased by $0.4~\mathrm{pu}$. The dual-port GFM control and DC-GFM control pass this imbalance on to the AC side (i.e., lowering the AC side frequency), while the AC-GFM control maintains the previous MMC power transfer and MMC AC  frequency. Next, at $t=0.3~\mathrm{s}$, the AC load is increased by $0.38~\mathrm{pu}$ and at $t=0.6~\mathrm{s}$ it is lowered by $0.26~\mathrm{pu}$. In response, the dual-port GFM control and AC-GFM control pass on this imbalance to the DC side (i.e., first lowering and then increasing the DC voltage), while the DC-GFM control maintains the previous MMC power transfer and MMC DC voltage. Moreover, it can be seen that, through this sequence of events, the single-port GFM controls control the total MMC energy to its nominal value, while its deviation is proportional to the power imbalance when using the dual-port GFM controls. At $t=1~\mathrm{s}$, the grid-forming AC source is redispatched to balance the system and all controls return the system to the nominal frequency and DC voltage. Finally, the voltage of the grid-forming AC source decreases by $0.1~\mathrm{pu}$ at $t=1.5~\mathrm{s}$ and at $t=2~\mathrm{s}$ the voltage of the grid-forming DC source decreases by $0.02~\mathrm{pu}$ and the resulting power imbalance is passed on to the AC side except when using the AC-GFM control.

After returning the nominal operating point at $t=2.5~\mathrm{s}$, we change the MMC power setpoint to $P^\star_{\text{MMC}}=P^\star_{\text{ac}}=P^\star_{\text{dc}}=-0.2~\mathrm{pu}$ at $t=3~\mathrm{s}$ to test the ability to control the power flow through the MMC, e.g., during emergency operation. Because the new power setpoints do not correspond to a steady-state, the controls cannot track the updated setpoint. Next, at $t=3.5~\mathrm{s}$ we update the MMC frequency and DC voltage setpoint to correspond to a steady state with $P^\star_{\text{MMC}}=-0.2~\mathrm{pu}$ and all controls track the setpoints. 

Finally, we return the system to the nominal operating point at $t=4~\mathrm{s}$ and disconnect the (grid-forming) DC source at $t=4.5~\mathrm{s}$. As expected, the DC-GFM control remains stable while the AC-GFM control fails to stabilize the system. At $t=5~\mathrm{s}$ we reconnect the AC source and reinitialize the simulation of the AC-GFM control and at $t=5.5~\mathrm{s}$ we disconnect the grid-forming AC source. As expected, the AC-GFM control remains stable while the DC-GFM control fails to stabilize the system. Notably, as predicted by the analysis in Sec.~\ref{sec:stability}, the two dual-port GFM controls maintain stable operation in both scenarios.

Moreover, comparing the dual-port GFM controls it can be seen that hybrid droop control results in significant oscillations and steady-state deviations when either the grid-forming AC or DC source are disconnected. We emphasize that the large steady-state deviations arise due to the active power and energy droop terms acting in the same direction and can be reduced by reducing the corresponding gains. However, in this case, pronounced oscillations occur during all transients due to the lower droop gain. In contrast, the energy-balancing control gains can be tuned separately to adjust the steady-state response and, with the exception of a large but brief overshoot, the response of energy-balancing control to disconnecting the DC source matches that of the DC-GFM control and the response to disconnecting the grid-forming AC source matches that of AC-GFM control.
\subsection{Case study: grid of grids}\label{sec:complextest}
Next, we consider the grid of grids shown in Fig.~\ref{fig:gogs} containing six $1000~\mathrm{MW}$ MMCs. At the rated power the MMC submodules provide $24.57~\mathrm{ms}$ of energy storage.

\begin{table}[b!]
\begin{center}
\caption{Control parameters: grid of grids\label{table:ctrparcomp}}
\addtolength{\tabcolsep}{-4pt}    
 \begin{tabular}{l| c c c c c c c}
 Control & $k^\text{ac}_p$\,[pu] & $k^\text{ac}_q$\,[pu] & $k^\text{dc}_p$\,[pu] & $k^\text{ac}_w$\,[pu] & $k^\text{dc}_w$\,[pu] & $\tau^\text{ac}$\,[s] & $\tau^\text{dc}$\,[s]\\ [0.5ex] 
\hline
hybrid droop & $0.05$ &  $0.05$ & $0.08125$ & $0.4$ & $0.4$ & \textendash & $0.001$ \\
energy-balan. & $0.01228$ &  $0.05$ & $0.004$ & $0.4$ & $0.4$ & $0.001$ & $0.001$ \\
\end{tabular}
\end{center}
\end{table}

\begin{figure*}
	\centering
	\includegraphics[width=1\textwidth]{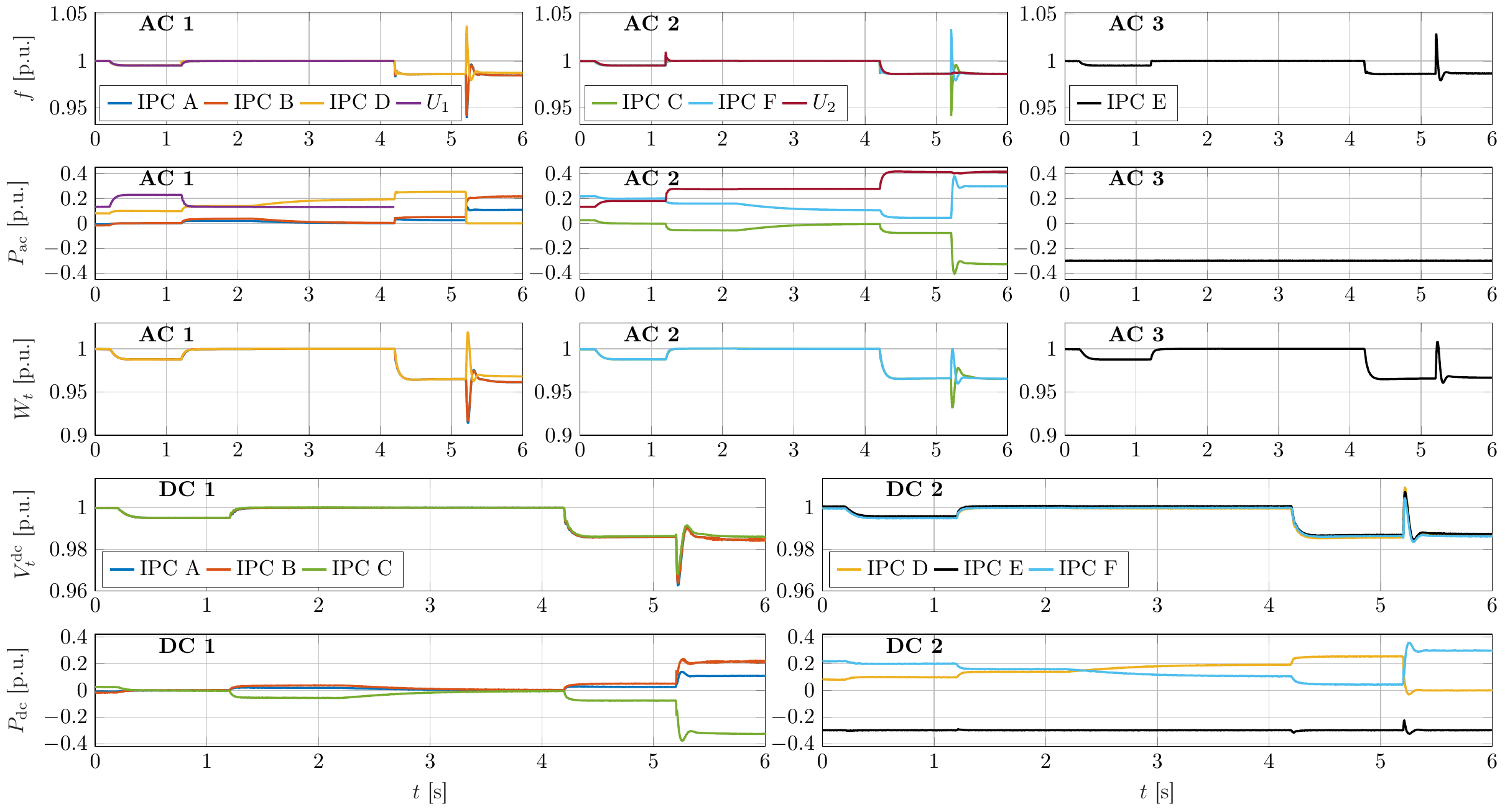}
		\caption{Response of energy balancing control to load changes and contingencies in the grid of grids test system shown in Fig.~\ref{fig:gogs}. \label{fig:MMCcomp}}
		\vspace{1em}
		\includegraphics[width=1\textwidth]{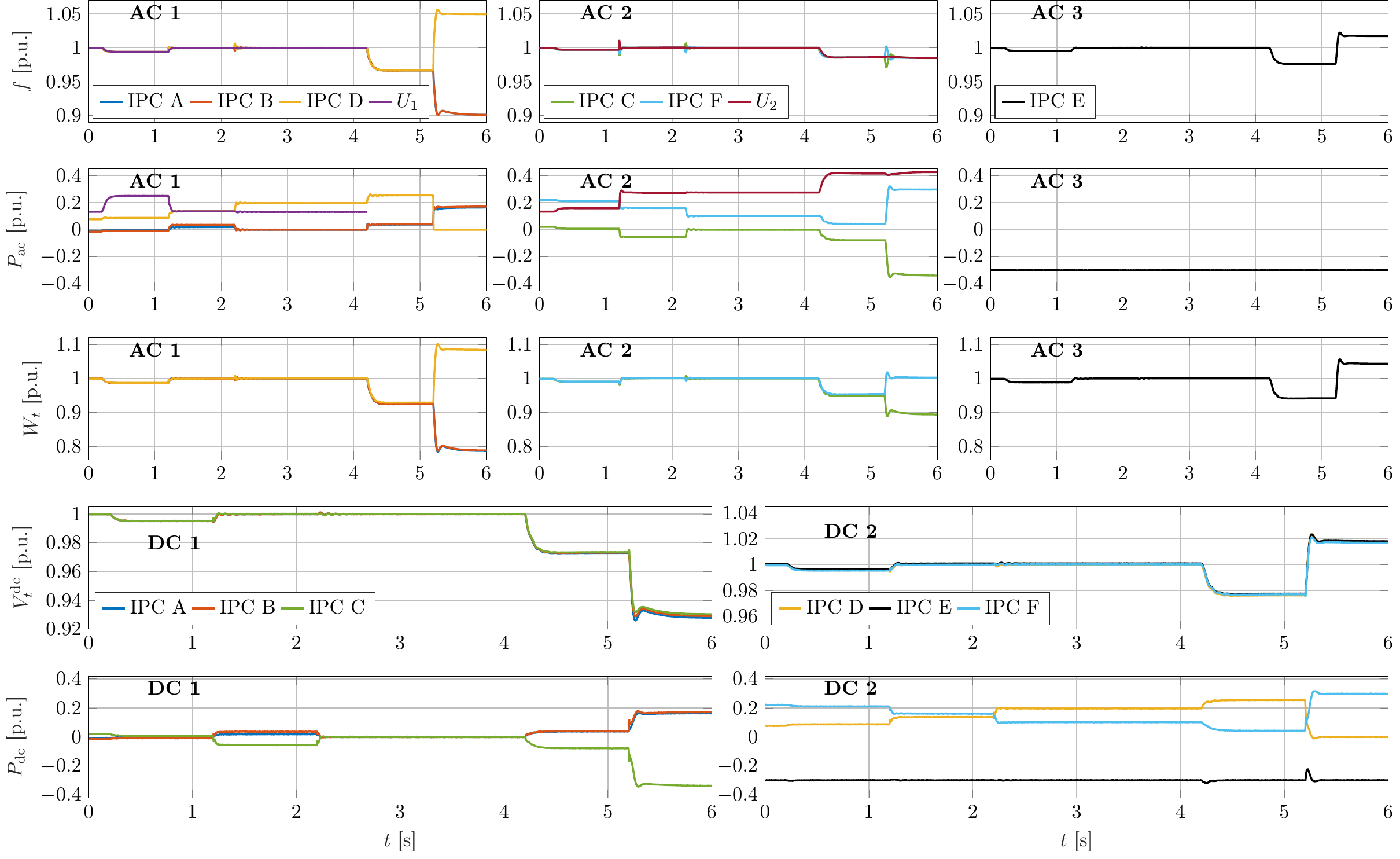}
	\caption{Response of hybrid droop control to load changes and contingencies in the grid of grids test system shown in Fig.~\ref{fig:gogs}. \label{fig:MMCcomp_ps}}
\end{figure*}

The system contains $P_{\text{ac}}$-$f$ droop controlled voltage sources $U_1$ and $U_2$ in the subgrids AC~1 and AC~2, and one PLL-based AC-GFL source in each AC subsystem. We note that AC~3 only contains grid-following generation and is formed by IPC E. Moreover, the MMCs are modeled using the averaged arm model, AC lines are modeled as $\pi$-line segments, and three parallel $\pi$ branches are used for DC lines. The initial dispatch and base values are shown in Fig.~\ref{fig:gogs}. Fig.~\ref{fig:MMCcomp} shows simulation results with energy balancing control for all IPCs and Fig.~\ref{fig:MMCcomp_ps} shows simulation results with hybrid droop control. The corresponding control gains can be found in Table~\ref{table:ctrparcomp}.

At $t=0.2~\mathrm{s}$, the power injection of the windfarm at node 3 is reduced to $10~\mathrm{MW}$ and it can be seen that both $U_1$ and $U_2$ respond by increasing their power injection. At $t=1.2~\mathrm{s}$, the power setpoint of $U_2$ is increased to $275~\mathrm{MW}$ to balance load and generation. The power flows through the networks DC~1 and DC~2 change autonomously, i.e., without redispatch, and the system returns to the nominal frequency. At this operating point, power flows from AC~3 through DC~1 to AC~1 and AC~2, but also from AC~2 through DC~2 to AC~1. While hybrid droop control is again prone to oscillations, up until this point hybrid droop control and energy-balancing control have comparable performance. Next, to avoid transferring power from AC~3 to AC~1 through AC~2 and illustrate that the steady-state power flow can be controlled through the DC voltage setpoints for energy-balancing control, we update $V^{\text{dc}\star}_t$ and $P^\star_{\text{ac}}=P^\star_{\text{dc}}$ (only for hybrid droop control) to correspond to a steady state with no power flow through DC~1 at $t=2.2~\mathrm{s}$. For both controls, the DC voltages and DC powers converge to their setpoints while the AC systems remain at their nominal frequencies. Because hybrid droop control explicitly controls active power to its setpoint it results in faster convergence of the DC power to the new operating point. Next, we test the response to severe contingencies. At $t=4.2~\mathrm{s}$, the droop controlled voltage source $U_1$ is disconnected and the only remaining grid-forming generator is $U_2$. As a result of removing the power generation by $U_1$, the overall system is imbalanced, but the IPCs continue to form a stable system AC~1 and the droop controlled source $U_2$ responds by increasing its power injection. We note that the configuration with pre-assigned single-port GFM roles used in \cite{GSA+20} would not have survived disconnecting $U_1$. Finally, at $t=5.2~\mathrm{s}$, we disconnect the  AC side of IPC D and observe that, despite a large transient, the system remains stable. Moreover, this event illustrates the key difference between hybrid droop control and energy-balancing control: hybrid droop control tries to maintain the scheduled pre-contingency power flow while energy-balancing control rebalances the system irrespective of the deviation from the scheduled power flow. At the nominal operating point power flows from AC~3 through DC~2 to AC~1 and AC~2 and no power flows through DC~1. However, after disconnecting $U_1$ and the AC side of IPC D, power has to flow from AC~2 to AC~1 through DC~1 and the power setpoint used by hybrid droop control is no longer reachable. Nonetheless, the $P_{\text{ac}}$-$f$ droop term in hybrid droop control aims to keep the IPCs close to the nominal power flow and, consequently, results in a large frequency drop in AC~1 and a counter-intuitive frequency above the nominal frequency in AC~3. In contrast, energy-balancing control, as predicted in Sec.~\ref{sec:ctrspecs}, synchronizes the entire system, i.e., DC voltages and AC frequencies are below the nominal value in all subgrids. Notably, because no power generation was lost by disconnecting the AC side of IPC D, the AC frequency and DC voltage deviations do not change significantly using energy-balancing control.

\section{Discussion and comparison}\label{sec:disc}
IPCs using dual-port GFM control leverage the MMCs capabilities to form both its AC and DC terminal voltage and simultaneously control the MMCs internal energy through both terminals. As a result, grids of grids with IPCs using the dual-port GFM controls proposed in Sec.~\ref{sec:dualgridform} are more resilient to severe contingencies (i.e., line and generator outages) than single-port GFM controls. In addition to this feature, the controls differ in several other key areas. 

\subsection{Single-port GFM control}
A strength of the standard single-port GFM approach is that tuning its parameters both for steady-state and dynamic specifications is easy to understand. In particular, the impact of the control gains on the steady-state response to power imbalances is straightforward and can be used to determine the control gains. Moreover, because this approach uses proportional active power control with respect to its dispatch point we observe fast convergence to a new operating point after a redispatch. On the other hand, single-port GFM IPCs require careful assignment of the GFM and GFL roles \cite{GSA+20} and, as discussed in Sec.~\ref{sec:sosgridform:disc} and demonstrated in Sec.~\ref{sec:case_study:simp}, lacks robustness to contingencies such as generator or transmission line outages. Specifically, single-port AC-GFM (DC-GFM) fails if no grid-forming resource (e.g., generator or GFM IPC) is available on the DC grid (AC grid). Moreover, the PLL used in single-port DC-GFM (i.e., AC-GFL) control is vulnerable to well-documented stability issues \cite{NERC17} during contingencies.

\subsection{Dual-port GFM control}
In contrast, the proposed dual-port GFM controls (i.e., hybrid power/energy droop control \eqref{eq:nonstiffform} and energy-balancing control \eqref{eq:nopowersetpoint}) do not require pre-assigning roles and maintain grid stability as long as a grid-forming source is available in one of the AC or DC subgrids (see Sec.~\ref{sec:complextest}). As a result hybrid droop control is more resilient to contingencies and changes in system topology while retaining the proportional active power control of single-port GFM that results in fast convergence to a new operating point after a redispatch. {However, if the IPCs' power setpoints are no longer reachable after a contingency, the energy and power droop terms in \eqref{eq:nonstiffform} represent conflicting objectives. This can result in large and counter-intuitive steady-state frequency deviations that have a complex dependence on the control gains and system-wide power flows (see Sec.~\ref{sec:ctrspecs}).} As a result, satisfying both steady-state and dynamic control specifications may be difficult (see Sec.~\ref{sec:case_study:simp}). In contrast to hybrid droop control, the control gains of energy-balancing control have a straightforward interpretation (see Sec.~\ref{sec:ctrspecs}) and allow separate tuning of the steady-state and dynamic response. Notably, energy-balancing control can match the dynamic and steady-state performance of single-port GFM control in our case studies (see Sec.~\ref{sec:case_study:simp}). 

\subsection{Dual-port GFM control without power setpoints}
Energy-balancing control \eqref{eq:nopowersetpoint} does not aim to control the power injection to a fixed setpoint on either terminal of the MMC and relies on the DC voltage and frequency setpoints for dispatch. As a result the speed of convergence to a new operating point after a redispatch is limited by the $l/r$ ratio of the DC cable and can be relatively slow compared to single-port GFM and hybrid droop control. Moreover, we expect that the dispatch of energy-balancing control may be sensitive to inaccurate DC cable parameters.

On the other hand, not relying on explicit power control increases the resilience of the IPCs and overall system to contingencies. In particular, energy-balancing control will not attempt to keep the IPCs close to an infeasible operating point but autonomously rebalance the power flows across the grid of grids. Moreover, not requiring a dispatch of power setpoints may be a significant advantage of  \eqref{eq:nopowersetpoint} in application scenarios with communication constraints or very large number of devices.

\subsection{Current limits and short-circuit faults}
It is well known that standard current reference limiting approaches (i.e., limiting the reference provided to the inner current controls) typically used in GFL control of VSCs (see e.g., \cite{PD15}) are incompatible with AC-GFM control and may result in instability of both single-port GFM and dual-port GFM controls. For two-level VSCs several heuristics (e.g., threshold virtual impedance \cite{PD15}) have been proposed that can limit the AC grid current without loss of stability in some limited scenarios (e.g., balanced short-circuit faults). However, the methods for two-level VSCs leverage the fact that limiting the AC grid current directly limits the current across the semiconductor switches. This is not true for MMCs. While fault control schemes for MMCs in AC-GFL mode have received some attention in the literature (see e.g., \cite{SAM+21}), the problem of how to limit MMC arm currents when one or two MMC terminals are in GFM mode and overload or a short-circuit fault occurs is largely open and requires significant future research.

\section{Conclusions and Outlook}\label{sec:conclusions}
This paper considered future power systems in which multiple non-synchronous HVAC systems are interconnected by Interconnecting Power Converters (IPC) and (multi-terminal) HVDC systems. In this context, we explored Modular Multilevel Converter (MMC) control and introduced the concept of dual-port grid-forming (GFM) control. In contrast to state-of-the-art single-port GFM control, dual-port GFM leverages the MMCs capabilities to simultaneously form both its AC and DC terminal voltage. Two implementations of the dual-port GFM control concept are presented and the differences between single-port GFM and dual-port GFM controls are discussed in depth. High-fidelity simulations are used to (i) illustrate and evaluate the approaches, and to (ii) demonstrate the resilience of grids of grids to severe contingencies (i.e., line and generator outages) when using dual-port GFM control. Future work will study the dynamic interactions of single-port GFM IPCs, dual-port GFM IPCs, and synchronous machines, {the inherent inertia response of dual-port GFM control}, and address current limiting and short-circuit fault ride through for single-port GFM and dual-port GFM MMC control.
\bibliographystyle{IEEEtran}
\bibliography{IEEEabrv,bibliography}

\end{document}